# 4-D Epanechnikov Mixture Regression in Light Field Image Compression

Boning Liu, Yan Zhao*, *Member, IEEE,* Xiaomeng Jiang, Shigang Wang, *Member, IEEE,* and Jian Wei

*Abstract*—With the emergence of light field imaging in recent years, the compression of its elementary image array (EIA) has become a significant problem. Our coding framework includes modeling and reconstruction. For the modeling, the covariance-matrix form of the 4-D Epanechnikov kernel (4-D EK) and its correlated statistics were deduced to obtain the 4-D Epanechnikov mixture models (4-D EMMs). A 4-D Epanechnikov mixture regression (4-D EMR) was proposed based on this 4-D EK, and a 4-D adaptive model selection (4-D AMLS) algorithm was designed to realize the optimal modeling for a pseudo video sequence (PVS) of the extracted key-EIA. A linear function based reconstruction (LFBR) was proposed based on the correlation between adjacent elementary images (EIs). The decoded images realized a clear outline reconstruction and superior coding efficiency compared to high-efficiency video coding (HEVC) and JPEG 2000 below approximately 0.05 bpp. This work realized an unprecedented theoretical application by (1) proposing the 4-D Epanechnikov kernel theory, (2) exploiting the 4-D Epanechnikov mixture regression and its application in the modeling of the pseudo video sequence of light field images, (3) using 4-D adaptive model selection for the optimal number of models, and (4) employing a linear function-based reconstruction according to the content similarity.

## I. INTRODUCTION

LIGHT field (LF) imaging has progressively developed over the past twenty years, and may become the most promising visual technology because of its ability to capture high-dimensional visual information [1]. It can realize glass-off 3D display through its true volume spatial optical capture system [2]. Two forms of LF image representations are shown in Fig. 1. However, because an LF image has a large volume of information for transformation, the use of compression technology has been essential.

Transform-based algorithms are the most commonly used in image coding [3]. Some high-efficiency video coding (HEVC)-embedded methods have good coding efficiency [4][5]. For example, [6] had the goal of efficiently coding plenoptic captured contents without knowing any camera geometries.

Manuscript received February 26, 2021; revised July 10, 2021, and August 7, 2021; accepted August 9, 2021. This work is supported by the National Natural Science Foundation of China (No.61631009, No. 61771220, No.61901187) and the National Key R&D Program of China (2017YFB1002900, 2017YFB0404800). (*Corresponding Author: Yan Zhao*)
Boning Liu, Yan Zhao, Shigang Wang, and Jian Wei are with the College of Communication Engineering, Jilin University, Changchun, 130012, P. R. China (e-mail: liuboning_jlu@sina.com; zhao_y@jlu.edu.cn; wangshigang@vip.sina.com; weijian@jlu.edu.cn).
Xiaomeng Jiang is with the College of Mathematics, Jilin University, Changchun, 130012, P. R. China (e-mail: jxmlucy@sina.com).

Moreover, [7] used a quadtree-based 2-D hierarchical method for LF image data compression, and [8] proposed a framework that exploited the inter- and intra-view correlations of LF images through sub-aperture images (SAIs) to produce a better compression performance. In addition to the algorithms implemented under HEVC, the research on image coding has also expanded to kernel regression frameworks such as the *Steered Mixture-of-Experts* (SMoE), in which the *Gaussian Mixture Model* (GMM) was used to reconstruct 4-D and 5-D LF images [9][10]. The SMoE method is a totally modeling-based algorithm, which is a revolutionary compression concept not only in image coding [11], but also in video coding and representation [12]. However, the algorithm is limited in relation to the kernel type, which means it has greater exploration potential in mathematical theory. Therefore, the aim of this paper is to explore the new kernel theory and modeling application in LF image compression. In addition to drastically departing from the traditional framework and using only a modeling method for EIs, a brand new regression theory is exploited.

The *Epanechnikov Kernel* (EK) is an excellent kernel function because of its convenient computation and local distribution support [13]. A 1-D EK was recently used to estimate the distribution of future wind power through probability density forecasting [14]. A 2-D EK was used as a filtering kernel in the context of photon mapping density estimation [15]. In addition, the latest SVM- and SVM-PSO-based calibration methods for JPEG images have also utilized a 2-D EK [16]. The theory of higher dimensions for an EK and its applications has not previously been reported. Moreover, its computability and distribution centrality advantages are also good reasons to explore it.

Our previous work included a rudimentary study of image coding based on a 3-D Epanechnikov kernel with a matrix form [17]. The experimental results showed that the discontinuity made the EK an energy-concentrated function and produced efficient results compared with a Gaussian kernel. Subsequently, in [18], we proposed a coding framework using 3-D Epanechnikov mixture regression (EMR) and Gaussian mixture regression (GMR) that could outperform JPEG image compression. Different from our previous papers, the study in this study aims to show that the EI-correlation of an LF image can be utilized for compression. Furthermore, the mathematical framework of a four-dimensional (4-D) application of the EK can be exploited to represent a pseudo video sequence (PVS).

This paper utilizes the theory of mathematical statistics and kernel density estimation to provide a complete derivation of the 4-D EK and its related statistics. An accurate covariance-matrix expression of the 4-D EK is exploited, as well as its marginal distribution, conditional distribution, and conditional mean, which are necessary for the proposed *4-D Epanechnikov mixture regression* (4-D EMR). There are many similar EIs in the EI array (EIA) of an LF image. According to the offsets of adjacent image blocks, key-EIs are extracted at specific intervals and arranged into a PVS, which is then modeled by the proposed theory. Under the Bayesian framework, PVS is modeled by local experts using *4-D Epanechnikov mixture model*s (4-D EMMs) with global support, and then each frame can be reconstructed using parameters obtained from the 4-D EMR. Finally, other EIs are predicted from the references using the content similarity through our linear function based reconstruction (LFBR) algorithm. In summary, this framework is a totally modeling-based coding algorithm.

The novelties of this paper include the following: (1) the deduction of the covariance-matrix form of the 4-D EK and its correlated statistics; (2) a proposal of the 4-D EMR and its use in PVS modeling of the key-EIA; (3) application of 4-D adaptive model selection (4-D AMLS) for the optimal modeling; and (4) utilization of the LFBR algorithm according to the content similarity.

The rest of the paper is organized as follows. The next section shows the related works for LF image coding. Section III deduces the covariance-matrix form of the 4-D EK and its correlated statistics for the 4-D EMM. Section IV presents the 4-D EMR algorithm and the theoretical visualization. Then, the modeling effect of the 4-D EMR compared with 4-D GMR is shown in Section V. Next, Section VI proposes our coding framework, including the 4-D AMLS and LFBR, along with the corresponding coding details. Experimental results are shown in Section VII and the conclusion and future work are finally presented in Section VIII.

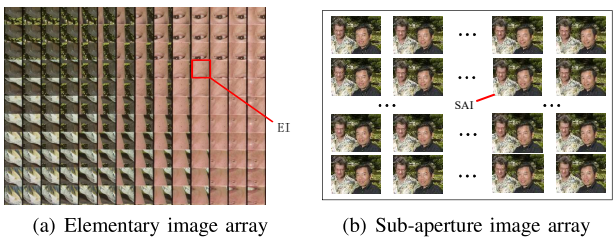

(a) Elementary image array    (b) Sub-aperture image array

Fig. 1. Two forms of LF image representation from *Sergio*. A sub-aperture image (SAI) is generated by selecting one pixel at the same horizontal and vertical location within each elemental image (EI).

## II. Related Works

There are a number of main approaches of the compression methods for LF images in literature: transform-based coding, prediction-based coding, view synthesis-based coding [19], and modeling-based coding.

As for the transform-based coding, the transformation can effectively extract the effective information of LF images in the transform domain. The LF codec in [20] fully exploited the 4D redundancy of LFs using a 4D-transform and hexadeca-trees. The 4D DCT-based technique was also included in JPEG Pleno [21]. Apart from DCT, an LF image compression scheme for sub-image-transformed elemental images using the Karhunen-Loeve transform (KLT) was proposed, which could efficiently utilize the similarity between EIs [22]. As for the 4D DWT, [23] proposed disparity-compensated lifting for the wavelet compression of LFs, in which the benefits of wavelet coding were obtained, such as scalable dimensions and superior compression performance. In [24], the authors proposed an LF codec with disparity guided sparse coding over a learned perspective-shifted LF dictionary. Another sparse coding algorithm was used to realize sparse reconstructions of 4D LFs from optimized 2-D projections using LF atoms as the fundamental building blocks [25]. These compression methods are based on transform, which are most typically used when addressing a growing coding problem.

In regard to prediction-based methods, [26] proposed a pseudo-sequence-based 2-D hierarchical coding structure for LF image compression. In [27], LF image data were decomposed into multiple views and used as a snake order pseudo-sequence input for HEVC. A hybrid scan order was also proposed in [28]. Based on inter-view prediction, [29] explored the LF structure, which could improve the behavior of pseudo-sequence-based lenslet image compression through sub-view rearrangement, illumination compensation, and reconstruction filtering. Moreover, [30] first designed a new LF-MVC prediction structure for LF images by extending the inter-view prediction into a two-directional parallel structure. In [31], the raw image was captured by an LF camera and decomposed into multiple views, which then constituted a pseudo sequence like video, so that the redundancy between views could be exploited by a video encoder. These studies considered the characteristic of LF image, in which EIs or SAIs are always regarded as PVS for coding.

There are many view synthesis-related algorithms, which may contain hybrid techniques for LF coding. Some of these methods include depth image-based rendering (DIBR). For example, [32] proposed to compress the LF image using a multi-view video plus depth (MVD) coding architecture, with an optimal slope decision algorithm designed to achieve a depth value. In [33], the authors introduced dubbed WaSPR (warping and sparse prediction on regions), which has additional features and improved performance compared to the WaSP (warping and sparse prediction) codec. Learning- and transform-based methods also constitute a large proportion of the available methods. The solution proposed in [34] was based on a graph learning approach to estimate the disparity among the views that compose the LF. [35] efficiently constructed the LF in the Fourier domain from a sparse set of views, which was then used to reconstruct intermediate viewpoints without requiring a disparity map. Deep learning has also been used to exploit the intrinsic similarity of LF images in a compression framework [36]. In [37], an LF image compression framework driven by a generative adversarial network (GAN) was proposed. There was also a new surface LF (SLF) representation based on the explicit geometry and a method for surface LF compression [38]. Overall, a more rational hybrid of techniques can make better use of the correlations within an LF image.

In contrast to these frameworks, Verhack proposed a completely modeling-based method for LF image coding in [11],

which has GMR at its core. This concept was a breakthrough for LF compression and inspired us to exploit more regression theories for this task. Therefore, in this paper, both the higher-dimensional EK theories and the brand-new framework on LF images are proposed. As for the LF compression framework, as mentioned in the literature above, the transform-based algorithms are the most common. To substitute the traditional transform-based idea with the proposed modeling-based method, we follow the PVS-based technique to model the key-EIs, and also combine with the similarity between EIs to reconstruct the unencoded EIs from the encoded key-EIs. Above all, this paper presents not only a mathematical advance to EK theory but also a new idea for LF image compression.

## III. MIXTURE-OF-EXPERTS REGRESSION WITH 4-D EPANECHNIKOV KERNEL

Under a Bayesian framework, it is assumed that the pixels of a PVS are distributed with global support with 4-D EKs, and then variables $x, y, z$, and $w$ are respectively defined as the abscissa value, ordinate value, time, and the gray value of each pixel in a PVS, respectively. In addition, $X, Y, Z$, and $W$ are the random values of $x, y, z$, and $w$, respectively. In terms of the formula expression, $\boldsymbol{\varphi} = (x, y, z, w)^T$ and $\boldsymbol{\delta} = (x, y, z)^T$. Therefore, the joint probability $p(\boldsymbol{\varphi})$ is given by the following formula with $K$ experts:

$$p(\boldsymbol{\varphi}) = \sum_{j=1}^{K} \alpha_j f_{\boldsymbol{\mu}_j, \boldsymbol{\Sigma}_j}(\boldsymbol{\varphi}), \quad (1)$$

where $f_{\boldsymbol{\mu}_j, \boldsymbol{\Sigma}_j}(\boldsymbol{\varphi})$ is the 4-D EK, $\alpha_j$ is the prior, and $\sum_{j=1}^{K} \alpha_j = 1$. Besides, mean value of variable is $\boldsymbol{\mu}_j = (\mu_{X_j}, \mu_{Y_j}, \mu_{Z_j}, \mu_{W_j})^T$, within which $\hat{\boldsymbol{\mu}}_j = (\mu_{X_j}, \mu_{Y_j}, \mu_{Z_j})^T$.

$$\boldsymbol{\Sigma}_j = \begin{bmatrix} \Sigma_{X_j X_j} & \Sigma_{X_j Y_j} & \Sigma_{X_j Z_j} & \Sigma_{X_j W_j} \\ \Sigma_{Y_j X_j} & \Sigma_{Y_j Y_j} & \Sigma_{Y_j Z_j} & \Sigma_{Y_j W_j} \\ \Sigma_{Z_j X_j} & \Sigma_{Z_j Y_j} & \Sigma_{Z_j Z_j} & \Sigma_{Z_j W_j} \\ \Sigma_{W_j X_j} & \Sigma_{W_j Y_j} & \Sigma_{W_j Z_j} & \Sigma_{W_j W_j} \end{bmatrix}, \text{ including } \boldsymbol{R}_j =$$

$$\begin{bmatrix} \Sigma_{X_j X_j} & \Sigma_{X_j Y_j} & \Sigma_{X_j Z_j} \\ \Sigma_{Y_j X_j} & \Sigma_{Y_j Y_j} & \Sigma_{Y_j Z_j} \\ \Sigma_{Z_j X_j} & \Sigma_{Z_j Y_j} & \Sigma_{Z_j Z_j} \end{bmatrix}.$$

Our reconstruction is based on (2) using the parameter set $\Omega_j = (\alpha_j, \boldsymbol{\mu}_j, \boldsymbol{\Sigma}_j)$ according to the *Mixture-of-Experts* (ME) algorithm which has been widely applied for linear regression, classification, and the mixtures of Gaussian processes [39]. The gray value of the modeled image $\widetilde{m}(\boldsymbol{\delta})$ for each location $\boldsymbol{\delta}$ is obtained from the conditional mean $m_{\boldsymbol{\mu}_j, \boldsymbol{\Sigma}_j}(\boldsymbol{\delta})$ and its gate function $g_{\hat{\boldsymbol{\mu}}_j, \boldsymbol{R}_j, \alpha_j}(\boldsymbol{\delta})$, which indicates the importance of each model within the global support:

$$\widetilde{m}(\boldsymbol{\delta}) = \sum_{j=1}^{K} g_{\hat{\boldsymbol{\mu}}_j, \boldsymbol{R}_j, \alpha_j}(\boldsymbol{\delta}) m_{\boldsymbol{\mu}_j, \boldsymbol{\Sigma}_j}(\boldsymbol{\delta}), \quad (2)$$

The function of the 4-D EK is calculated in Section III-A. The gate function and the conditional mean are deduced in Section III-B and Section III-C, respectively.

### A. 4-D EK

We determine the expression for the 4-D EK $f_{\boldsymbol{\mu}_j, \boldsymbol{\Sigma}_j}(\boldsymbol{\varphi})$ according to the principle that a kernel function $K(\boldsymbol{x})$ must satisfy $\int K(\boldsymbol{x})d\boldsymbol{x} = 1$ and $K(\boldsymbol{x}) \geqslant 0$ [40]. To simplify the calculation, we can firstly consider a basic form of the 4-D EK with sample $\hat{\boldsymbol{\varphi}} = (\hat{x}, \hat{y}, \hat{z}, \hat{w})^T$ and $\boldsymbol{\Lambda} = \mathbf{diag}(\frac{1}{a}, \frac{1}{b}, \frac{1}{c}, \frac{1}{d})$, where $a, b, c, d$ are the corresponding ellipsoid axial lengths:

$$\hat{f}(\hat{\boldsymbol{\varphi}}) = \begin{cases} k\left(1 - \hat{\boldsymbol{\varphi}}^T \boldsymbol{\Lambda}^2 \hat{\boldsymbol{\varphi}}\right), & \hat{\boldsymbol{\varphi}}^T \boldsymbol{\Lambda}^2 \hat{\boldsymbol{\varphi}} \leq 1 \\ 0, & \text{otherwise}, \end{cases} \quad (3)$$

in which $k$ is the normalized coefficient.

$\hat{f}(\hat{\boldsymbol{\varphi}})$ is the probability density function whose global integral is 1; therefore, the coefficient $k$ satisfies

$$k \iiiint_{\Omega_{\hat{\boldsymbol{\varphi}}}} \left(1 - \hat{\boldsymbol{\varphi}}^T \boldsymbol{\Lambda}^2 \hat{\boldsymbol{\varphi}}\right) d\Omega_{\hat{\boldsymbol{\varphi}}} = 1, \ \left\{\Omega_{\hat{\boldsymbol{\varphi}}} : \hat{\boldsymbol{\varphi}}^T \boldsymbol{\Lambda}^2 \hat{\boldsymbol{\varphi}} \leq 1\right\}. \quad (4)$$

From Appendix A, we can determine the integral result of $\iiiint_{\Omega_{\hat{\boldsymbol{\varphi}}}} \left(1 - \hat{\boldsymbol{\varphi}}^T \boldsymbol{\Lambda}^2 \hat{\boldsymbol{\varphi}}\right) d\Omega_{\hat{\boldsymbol{\varphi}}}$ is $\pi^2 / (6|\boldsymbol{\Lambda}|)$. Thus, we can know

$$k = \frac{6|\boldsymbol{\Lambda}|}{\pi^2}. \quad (5)$$

Thus far, the basic form of the 4-D EK is

$$\hat{f}(\hat{\boldsymbol{\varphi}}) = \begin{cases} \frac{6|\boldsymbol{\Lambda}|}{\pi^2}\left(1 - \hat{\boldsymbol{\varphi}}^T \boldsymbol{\Lambda}^2 \hat{\boldsymbol{\varphi}}\right), & \hat{\boldsymbol{\varphi}}^T \boldsymbol{\Lambda}^2 \hat{\boldsymbol{\varphi}} \leq 1 \\ 0, & \text{otherwise}. \end{cases} \quad (6)$$

Thereafter, we determine the covariance matrix $\hat{\boldsymbol{\Sigma}}$ of $\hat{f}(\hat{\boldsymbol{\varphi}})$ whose detailed verification is provided in Appendix B.

$$\hat{\boldsymbol{\Sigma}} = \frac{1}{8}\left(\boldsymbol{\Lambda}^{-1}\right)^2, \quad \boldsymbol{\Lambda}^2 = \frac{1}{8}\hat{\boldsymbol{\Sigma}}^{-1}. \quad (7)$$

As a result, the function with respect to the covariance matrix $\hat{\boldsymbol{\Sigma}}$ is $\hat{f}(\hat{\boldsymbol{\varphi}})$:

$$\hat{f}(\hat{\boldsymbol{\varphi}}) = \begin{cases} \frac{6|\boldsymbol{\Lambda}|}{\pi^2}\left(1 - \hat{\boldsymbol{\varphi}}^T \cdot \frac{1}{8}\hat{\boldsymbol{\Sigma}}^{-1} \cdot \hat{\boldsymbol{\varphi}}\right), & \frac{1}{8}\hat{\boldsymbol{\varphi}}^T \hat{\boldsymbol{\Sigma}}^{-1} \hat{\boldsymbol{\varphi}} \leq 1 \\ 0 & \text{otherwise}. \end{cases} \quad (8)$$

We implement a linear transformation to the $\boldsymbol{\varphi}$ domain including translation and rotation with $\boldsymbol{\mu}_j = (\mu_{X_j}, \mu_{Y_j}, \mu_{Z_j}, \mu_{W_j})^T$ and $\boldsymbol{U}$. The term $\boldsymbol{U}$ is an orthonormal matrix that satisfies $\boldsymbol{U}^{-1} = \boldsymbol{U}^T$ and $\det(\boldsymbol{U}) = 1$.

$$\hat{\boldsymbol{\varphi}} = \boldsymbol{U} \cdot (\boldsymbol{\varphi} - \boldsymbol{\mu}_j). \quad (9)$$

Substituting (9) into (6) results in

$$\hat{f}(\boldsymbol{U}(\boldsymbol{\varphi} - \boldsymbol{\mu}_j))$$
$$= \begin{cases} \frac{6|\boldsymbol{\Lambda}|}{\pi^2}\left[1 - (\boldsymbol{\varphi} - \boldsymbol{\mu}_j)^T \boldsymbol{U}^{-1} \boldsymbol{\Lambda}^2 \boldsymbol{U}(\boldsymbol{\varphi} - \boldsymbol{\mu}_j)\right], & \boldsymbol{\Omega}_{\hat{f}(\boldsymbol{U}(\boldsymbol{\varphi} - \boldsymbol{\mu}_j))} \\ 0, & \text{otherwise}, \end{cases} \quad (10)$$

in which

$$\boldsymbol{\Omega}_{\hat{f}(\boldsymbol{U}(\boldsymbol{\varphi} - \boldsymbol{\mu}_j))} : (\boldsymbol{\varphi} - \boldsymbol{\mu}_j)^T \boldsymbol{U}^{-1} \boldsymbol{\Lambda}^2 \boldsymbol{U}(\boldsymbol{\varphi} - \boldsymbol{\mu}_j) \leq 1. \quad (11)$$

Combined with (8) and (10), the general form of covariance matrix $\boldsymbol{\Sigma}_j$ is formulated as

$$\frac{1}{8}\boldsymbol{\Sigma}_j^{-1} = \boldsymbol{U}^{-1}\boldsymbol{\Lambda}^2 \boldsymbol{U}. \quad (12)$$

Therefore, the determinant of $\boldsymbol{\Lambda}$ is

$$|\boldsymbol{\Lambda}| = \frac{1}{64\sqrt{|\boldsymbol{\Sigma}_j|}}. \quad (13)$$

Substituting (12) and (13) into (10), we obtain the general form of the 4-D EK $f_{\mu_j,\Sigma_j}(\varphi)$ respect to $\mu_j$ and $\Sigma_j$ as

$$f_{\mu_j,\Sigma_j}(\varphi)=\begin{cases}\frac{3}{32\pi^2\sqrt{|\Sigma_j|}}[1-\frac{1}{8}(\varphi-\mu_j)^T\Sigma_j^{-1}(\varphi-\mu_j)], & \Omega_{f_{\mu_j,\Sigma_j}(\varphi)}\\ 0 & \text{otherwise.}\end{cases} \quad (14)$$

in which

$$\Omega_{f_{\mu_j,\Sigma_j}(\varphi)}:\frac{1}{8}(\varphi-\mu_j)^T\Sigma_j^{-1}(\varphi-\mu_j)\leq 1. \quad (15)$$

*B. Gate function of 4-D EMM*

The gate function is given by $g_{\hat{\mu}_j,R_j,\alpha_j}(\delta)$

$$g_{\hat{\mu}_j,R_j,\alpha_j}(\delta)=\frac{\alpha_j F_{\hat{\mu}_j,R_j}(\delta)}{\sum_{j=1}^K \alpha_j F_{\hat{\mu}_j,R_j}(\delta)}. \quad (16)$$

where $F_{\hat{\mu}_j,R_j}(\delta)$ is the marginal distribution of location information. In this modeling, the gray values of different positions are independent of each other; therefore,

$$F_{\hat{\mu}_j,R_j}(\delta)=\int f_{\mu_j,\Sigma_j}(\varphi)dw. \quad (17)$$

For the simplicity of calculation, we define

$$\varphi'=(x',y',z',w')^T=\varphi-\mu_j. \quad (18)$$

$$\delta'=(x',y',z')^T=\delta-\hat{\mu}_j. \quad (19)$$

Thus, the 4-D EK function becomes

$$f_{\mu_j=0,\Sigma_j}(\varphi')=\begin{cases}\frac{3}{32\pi^2\sqrt{|\Sigma_j|}}(1-\frac{1}{8}\varphi'^T\Sigma_j^{-1}\varphi'), & \frac{1}{8}\varphi'^T\Sigma_j^{-1}\varphi'\leq 1\\ 0 & \text{otherwise.}\end{cases} \quad (20)$$

According to (17), we can obtain

$$F_{\hat{\mu}_j=0,R_j}(\delta')=\int f_{\mu_j=0,\Sigma_j}(\varphi')dw'$$
$$=\int_{w_1'}^{w_2'}\frac{3\left(1-\frac{1}{8}\varphi'^T\Sigma_j^{-1}\varphi'\right)}{32\pi^2\sqrt{|\Sigma_j|}}dw'. \quad (21)$$

To solve the integral, the parameters are set as follows:

$$\Sigma_j^{-1}=\begin{bmatrix}\Sigma_{X_jX_j} & \Sigma_{X_jY_j} & \Sigma_{X_jZ_j} & \Sigma_{X_jW_j}\\ \Sigma_{Y_jX_j} & \Sigma_{Y_jY_j} & \Sigma_{Y_jZ_j} & \Sigma_{Y_jW_j}\\ \Sigma_{Z_jX_j} & \Sigma_{Z_jY_j} & \Sigma_{Z_jZ_j} & \Sigma_{Z_jW_j}\\ \Sigma_{W_jX_j} & \Sigma_{W_jY_j} & \Sigma_{W_jZ_j} & \Sigma_{W_jW_j}\end{bmatrix}^{-1}=\begin{bmatrix}\sigma_a & \sigma_b & \sigma_c & \sigma_d\\ \sigma_b & \sigma_e & \sigma_f & \sigma_g\\ \sigma_c & \sigma_f & \sigma_h & \sigma_i\\ \sigma_d & \sigma_g & \sigma_i & \sigma_j\end{bmatrix}. \quad (22)$$

In general, it is necessary to initially determine $w_1'$ and $w_2'$, which are the two endpoints of the range $\frac{1}{8}\varphi'^T\Sigma_j^{-1}\varphi'\leq 1$ and can be solved from

$$\frac{1}{8}(x',y',z',w')\Sigma_j^{-1}(x',y',z',w')^T-1=0, \quad (23)$$

$$\sigma_j w'^2+2(\sigma_d x'+\sigma_g y'+\sigma_i z')w'+[(\sigma_a x'+\sigma_b y'+\sigma_c z')x'\\ +(\sigma_b x'+\sigma_e y'+\sigma_f z')y'+(\sigma_c x'+\sigma_f y'+\sigma_h z')z'-8]=0. \quad (24)$$

Hence,

$$w_1'=\frac{-(\sigma_d x'+\sigma_g y'+\sigma_i z')-\sqrt{\Delta}}{\sigma_j}, w_2'=\frac{-(\sigma_d x'+\sigma_g y'+\sigma_i z')+\sqrt{\Delta}}{\sigma_j}, \quad (25)$$

in which

$$\Delta=(\sigma_d x'+\sigma_g y'+\sigma_i z')^2-[(\sigma_a x'+\sigma_b y'+\sigma_c z')x'+\\ (\sigma_b x'+\sigma_e y'+\sigma_f z')y'+(\sigma_c x'+\sigma_f y'+\sigma_h z')z'-8]\sigma_j. \quad (26)$$

Therefore, when $\frac{1}{8}\varphi'^T\Sigma_j^{-1}\varphi'\leq 1$, $F_{\hat{\mu}_j=0,R_j}(\delta')$ is

$$F_{\hat{\mu}_j=0,R_j}(\delta')=\frac{3\int_{w_1'}^{w_2'}\left(1-\frac{1}{8}\varphi'^T\Sigma_j^{-1}\varphi'\right)dw'}{32\pi^2\sqrt{|\Sigma_j|}}=\frac{3}{32\pi^2\sqrt{|\Sigma_j|}}\cdot\frac{\Delta^{\frac{3}{2}}}{6\sigma_j^2}$$

$$=\frac{3}{32\pi^2\sqrt{|\Sigma_j|}}\cdot\frac{|\Sigma_j|^{\frac{1}{2}}\left(8|R_j|-\delta'^T|R_j|R_j^{-1}\delta'\right)^{\frac{3}{2}}}{6|R_j|^2}$$

$$=\frac{\sqrt{2}}{4\pi^2\sqrt{|R_j|}}\cdot\left(1-\frac{1}{8}\delta'^T R_j^{-1}\delta'\right)^{\frac{3}{2}}. \quad (27)$$

Substitution of (19) back into (27) yields the general form of $F_{\hat{\mu}_j,R_j}(\delta)$:

$$F_{\hat{\mu}_j,R_j}(\delta)=\begin{cases}\frac{\sqrt{2}}{4\pi^2\sqrt{|R_j|}}\left[1-\frac{1}{8}(\delta-\hat{\mu}_j)^T R_j^{-1}(\delta-\hat{\mu}_j)\right]^{\frac{3}{2}}, & \Omega_{F_{\hat{\mu}_j,R_j}(\delta)}\\ 0, & \text{otherwise.}\end{cases} \quad (28)$$

in which

$$\Omega_{F_{\hat{\mu}_j,R_j}(\delta)}:\frac{1}{8}(\delta-\hat{\mu}_j)^T R_j^{-1}(\delta-\hat{\mu}_j)\leq 1 \quad (29)$$

*C. Conditional Mean of W|XYZ*

The gray values $W$ are independent of each other at different $(X,Y,Z)$. The conditional mean $m_{\mu_j,\Sigma_j}(\delta)$ must be solved through the conditional distribution of $W$ under $(X,Y,Z)$, which is defined as $e_{\mu_j,\Sigma_j}(w|\delta)$:

$$e_{\mu_j,\Sigma_j}(w|\delta)=\frac{f_{\mu_j,\Sigma_j}(\varphi)}{F_{\hat{\mu}_j,R_j}(\delta)}=\\ \begin{cases}\frac{3|R_j|^{\frac{1}{2}}[1-\frac{1}{8}(\varphi-\mu_j)^T\Sigma_j^{-1}(\varphi-\mu_j)]}{8\sqrt{2|\Sigma_j|}[1-\frac{1}{8}(\delta-\hat{\mu}_j)^T R_j^{-1}(\delta-\hat{\mu}_j)]^{\frac{3}{2}}}, & \frac{1}{8}(\varphi-\mu_j)^T\Sigma_j^{-1}(\varphi-\mu_j)\leq 1\\ 0, & \text{otherwise.}\end{cases} \quad (30)$$

Combined with (19) and (25), the integral is given by

$$m_{\mu_j=(0,0,0,\mu_{W_j})^T,\Sigma_j}(\delta')=\int we_{\mu_j,\Sigma_j}(w|\delta)dw\\
=\int(w'+\mu_{W_j})e_{\Sigma_j}(w'|x',y',z')dw'\\
=\mu_{W_j}+\int_{w_1'}^{w_2'}w'\frac{\frac{3\left[1-\frac{1}{8}(x',y',z',w')\Sigma_j^{-1}(x',y',z',w')^T\right]}{32\pi^2\sqrt{|\Sigma_j|}}}{\frac{3}{32\pi^2\sqrt{|\Sigma_j|}}\cdot\frac{\Delta^{\frac{3}{2}}}{6\sigma_j^2}}\\
=\mu_{W_j}+\frac{|M|x'-|S|y'+|N|z'}{|R|}, \quad (31)$$

where the sub-matrices of the covariance matrix $\Sigma_j$ are

$$M=\begin{bmatrix}\Sigma_{X_jY_j} & \Sigma_{X_jZ_j} & \Sigma_{X_jW_j}\\ \Sigma_{Y_jY_j} & \Sigma_{Y_jZ_j} & \Sigma_{Y_jW_j}\\ \Sigma_{Z_jY_j} & \Sigma_{Z_jZ_j} & \Sigma_{Z_jW_j}\end{bmatrix}, S=\begin{bmatrix}\Sigma_{X_jX_j} & \Sigma_{X_jZ_j} & \Sigma_{X_jW_j}\\ \Sigma_{Y_jX_j} & \Sigma_{Y_jZ_j} & \Sigma_{Y_jW_j}\\ \Sigma_{Z_jX_j} & \Sigma_{Z_jZ_j} & \Sigma_{Z_jW_j}\end{bmatrix}, \text{ and}$$

$$N = \begin{bmatrix} \Sigma_{X_j X_j} & \Sigma_{X_j Y_j} & \Sigma_{X_j W_j} \\ \Sigma_{Y_j X_j} & \Sigma_{Y_j Y_j} & \Sigma_{Y_j W_j} \\ \Sigma_{Z_j X_j} & \Sigma_{Z_j Y_j} & \Sigma_{Z_j W_j} \end{bmatrix}.$$

Simplifying the calculation yields the following result:

$$m_{\boldsymbol{\mu}_j, \boldsymbol{\Sigma}_j}(\boldsymbol{\delta}) = \boldsymbol{\mu}_{W_j} + (\Sigma_{W_j X_j}, \Sigma_{W_j Y_j}, \Sigma_{W_j Z_j})\boldsymbol{R}_j^{-1}(\boldsymbol{\delta} - \hat{\boldsymbol{\mu}}_j). \quad (32)$$

## IV. 4-D EMR REPRESENTATION AND VARIABLE VISUALIZATION

### A. Representation and Parameter Optimization of 4-D EMR

This modeling-based algorithm is in accordance with the reconstruction formula in (2), and the parameter set of the $t$th iteration with all $K$ experts is $\boldsymbol{\Omega}_t = \{\boldsymbol{\mu}_{jt}, \boldsymbol{\Sigma}_{jt}, \alpha_{jt}\}, j = 1, 2, ..., K$, which needs to be optimized and encoded.

Before parameter estimation, the Y, U, and V channels of a PVS are represented separately. For a certain input channel, the total number of pixels is $N$, and the $i$th pixel is denoted as $(x_i, y_i, z_i, \widetilde{m_0}(x_i, y_i, z_i))$, where $x_i$, $y_i$, $z_i$, and $\widetilde{m_0}(x_i, y_i, z_i)$ are the X-coordinate, Y-coordinate, time, and gray value, respectively. Therefore, the data of a certain channel can be represented as an $N \times 4$ matrix $data_{N \times 4}$, which is a preparation for the modeling and optimization. The pseudo-code of the optimal algorithm is presented in Algorithm 1.

First, *K-Means Clustering* is used to approximately allocate the samples of $data_{N \times 4}$ into $K$ groups [41]. Thus, the initialization parameters of the $K$ models of the 4-D EMM are obtained as follows: $\boldsymbol{\Omega}_1 = \{\boldsymbol{\mu}_{j1}, \boldsymbol{\Sigma}_{j1}, \alpha_{j1}\}, j = 1, 2, ..., K$.

To estimate a relatively optimal parameter set for the EMM, the parameter estimation of the GMM obtained using the *Expectation-Maximization* (EM) algorithm is considered as an approximate iteration reference [42]. The core of the E-step is the posterior probability,

$$Qt_{ij} = \frac{\alpha_j f_{\boldsymbol{\mu}_j, \boldsymbol{\Sigma}_j}(\boldsymbol{\varphi}_i)}{p(\boldsymbol{\varphi}_i)} \quad (33)$$

and the iteration of the M-step is shown in Steps 7-9.

---

**Algorithm 1** 4-D Epanechnikov Mixture Regression
---
1: Initialize $data_{N \times 4}$ using k-Means into $K$ clusters and obtain the initial parameter set $\boldsymbol{\Omega}_1$.
2: Substitute $\boldsymbol{\Omega}_1$ into (2) and obtain the initial reconstructed PVS $\widetilde{m_1}(x_i, y_i, z_i)$. Next, MSE(1) $= \frac{1}{N}\sum_{i=1}^{N}(\widetilde{m_1}(x_i, y_i, z_i) - \widetilde{m_0}(x_i, y_i, z_i))^2$.
3: **for** $t = 2 : 14$
4:   **E-step**: Update the posterior probability $Qt_{ij}$,
5:     $i = 1, 2, ..., N, j = 1, 2, ..., K$
6:   **for** $j = 1 : K$
7:     **M-step**: Update $\boldsymbol{\mu}_{jt} = \frac{\sum_{i=1}^{N} Qt_{ij}\boldsymbol{\varphi}_i}{\sum_{i=1}^{N} Qt_{ij}}$,
8:     Update $\boldsymbol{\Sigma}_{jt} = \frac{\sum_{i=1}^{N} Qt_{ij}(\boldsymbol{\varphi}_i - \boldsymbol{\mu}_{jt})^T(\boldsymbol{\varphi}_i - \boldsymbol{\mu}_{jt})}{\sum_{i=1}^{N} Qt_{ij}}$,
9:     Update $\alpha_{jt} = \frac{\sum_{i=1}^{N} Qt_{ij}}{N}$
10:  **end**
11:  Substitute $\boldsymbol{\Omega}_t$ into (2) to obtain $\widetilde{m_t}(x_i, y_i, z_i)$.
12:  MSE$(t) = \frac{1}{N}\sum_{i=1}^{N}(\widetilde{m_t}(x_i, y_i, z_i) - \widetilde{m_0}(x_i, y_i, z_i))^2$
13: **end**
14: $t_0$=find(MSE==min(MSE))
15: **Output**: Optimal parameters $\boldsymbol{\Omega}_{t_0}$

---

In contrast to the traditional EM algorithm, the regression rule for each modeling result is adapted using the *Mean Square Error* (MSE) [43]. From hundreds of experiments, it was found that $t = 14$ is the proper choice for iteration, because it can achieve an optimal balance of regression effect and running speed. A total of 13 iterations of the framework are used, and 14 results are obtained, including the initialization. In this algorithm, the MSE between $\widetilde{m_t}(x_i, y_i, z_i), t = 1, 2, ..., 14$ and the original sequence $\widetilde{m_0}(x_i, y_i, z_i)$ is calculated according to Step 12. Finally, the result corresponding to the minimum MSE is the actual output.

To ensure that the model performance results are fair, all of the comparison experiments with the GMR were implemented under the same framework of Algorithm 1 by replacing the $f_{\boldsymbol{\mu}_j, \boldsymbol{\Sigma}_j}(\boldsymbol{\varphi}_i)$ in (33) with the Gaussian kernel distribution.

### B. Visualization of Key Statistics

To determine the quality of the modeled result, the core is $\widetilde{m}(\boldsymbol{\delta})$ in (2), in which $m_{\boldsymbol{\mu}_j, \boldsymbol{\Sigma}_j}(\boldsymbol{\delta})$ in (32) and gate function $g_{\hat{\boldsymbol{\mu}}_j, \boldsymbol{R}_j, \alpha_j}(\boldsymbol{\delta})$ in (16) are the essential parts. For example, a $37 \times 37 \times 4$ PVS is modeled using an EMR with 32 experts, as shown in Fig. 2(a), and the two key statistics are visualized

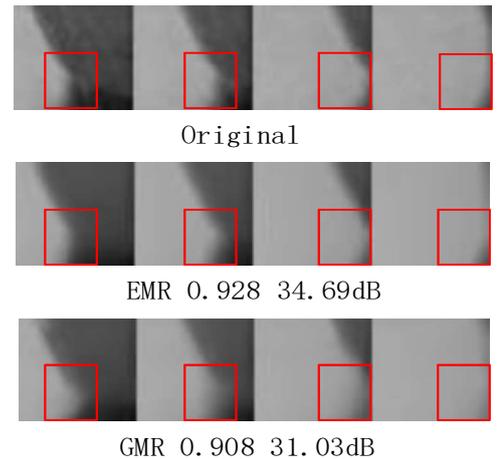

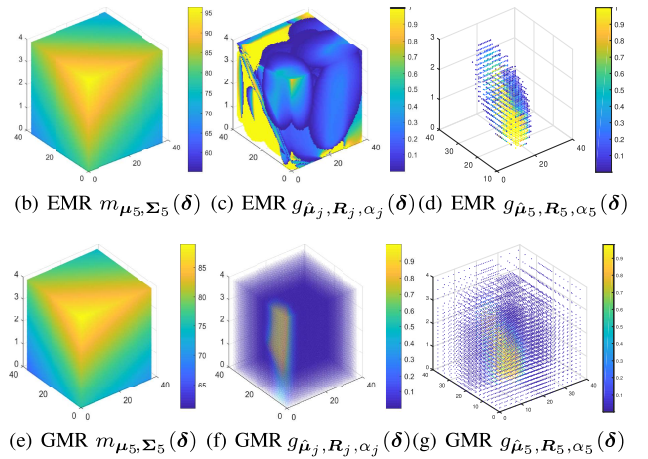

(a) Modeling results of $37 \times 37 \times 4$ pseudo sequence

(b) EMR $m_{\boldsymbol{\mu}_5, \boldsymbol{\Sigma}_5}(\boldsymbol{\delta})$  (c) EMR $g_{\hat{\boldsymbol{\mu}}_j, \boldsymbol{R}_j, \alpha_j}(\boldsymbol{\delta})$ (d) EMR $g_{\hat{\boldsymbol{\mu}}_5, \boldsymbol{R}_5, \alpha_5}(\boldsymbol{\delta})$

(e) GMR $m_{\boldsymbol{\mu}_5, \boldsymbol{\Sigma}_5}(\boldsymbol{\delta})$  (f) GMR $g_{\hat{\boldsymbol{\mu}}_j, \boldsymbol{R}_j, \alpha_j}(\boldsymbol{\delta})$ (g) GMR $g_{\hat{\boldsymbol{\mu}}_5, \boldsymbol{R}_5, \alpha_5}(\boldsymbol{\delta})$

Fig. 2. Experimental results modeled by EMR and GMR together with the regressing visualization. In (c) and (f), $j = 1, 2, 3, ..., 32$.

TABLE I
EXPERIMENTAL RESULTS OF MODELING A PVS USING EMR

| GOP | | GOP = 4 | | | GOP = 8 | | | GOP = 16 | | |
|---|---|---|---|---|---|---|---|---|---|---|
| Models | CB | 75 | 38 | 19 | 75 | 38 | 19 | 75 | 38 | 19 |
| 5120 | PSNR(dB) / SSIM | 26.30 / 0.856 | 26.56 / 0.851 | 27.17 / 0.846 | 25.10 / 0.832 | 25.80 / 0.835 | 26.73 / 0.843 | 23.97 / 0.802 | 24.88 / 0.817 | 24.19 / 0.788 |
| | T(s) | 217.44 | 39.22 | 7.20 | 740.40 | 116.01 | 19.65 | 1202.40 | 396.35 | 56.42 |

<sup>a</sup> 'Models' represents the total number of models for modeling the entire EIA.
<sup>b</sup> Time $T$ is drawn from the code running in MATLAB R2017b using Intel(R) Core(TM) i7-7700K CPU @ 4.20GHz.

in Fig. 2(b)-2(g). The modeling result of the EMR has a more distinct and accurate border than that of the GMR.

To observe a clearer distribution, the conditional mean of the 5th expert is visualized in the 4-D view shown in Fig. 2(b) and 2(e), in which the color represents the gray values in the $37 \times 37 \times 4$ location area. It can be observed that $m_{\boldsymbol{\mu}_5, \boldsymbol{\Sigma}_5}(\boldsymbol{\delta})$ is linear in a 3-D coordinate space.

The main difference between the 4-D EMR and 4-D GMR is the gate function. Regarding the observation $g_{\hat{\boldsymbol{\mu}}_j, \boldsymbol{R}_j, \alpha_j}(\boldsymbol{\delta}), j = 1, 2, ..., 32$ of the EMR in Fig. 2(c), the gate distribution of each kernel only offers support within an ellipsoid whose centroid coordinates are $\hat{\boldsymbol{\mu}}_j, j = 1, 2, ..., 32$. It can be determined from the color changes that the probabilities of each $g_{\hat{\boldsymbol{\mu}}_j, \boldsymbol{R}_j, \alpha_j}(\boldsymbol{\delta}), j = 1, 2, ..., 32$ become weaker until reaching a value of zero from the inside to the outside. However, $g_{\hat{\boldsymbol{\mu}}_j, \boldsymbol{R}_j, \alpha_j}(\boldsymbol{\delta}), j = 1, 2, ..., 32$ of the GMR are all continuous and flat with global support. Therefore, the gate functions can hardly be seen in Fig. 2(f). For better visualization, consider the gate function of the 5th kernel in Fig. 2(d) and 2(g). It is evident that the EMR simulates local support, whereas the GMR simulates global support. Both distributions decrease from the center of the kernel.

Therefore, EMR can outperform GMR for sequence modeling in some ways. The local support of the gate function of EMR makes the distribution function more compact and concrete, which leads to a more focused modeling effect.

## V. 4-D EMR MODELING EXPERIMENTS

### A. Motivation of the modeling comparison

These experiments were conducted in preparation for the compression framework in Section VI. The proposed method is a totally modeling-based algorithm consisting of EMR and GMR. Therefore, the effects of different kernels on various block sizes and image channels of the EI pseudo-sequence have to be exploited.

This subsection analyzes the 4-D modeling of the PVS of the selected EIs from the original EIA using the Epanechnikov kernel compared with the Gaussian kernel in experiments based on different sizes of *Coding Blocks* (CBs) and *Groups of Pictures* (GOPs). Only the modeling results of the key-EIA are shown without reconstructing the entire EIA, because this section only discusses the modeling behaviors of the two kernels.

### B. Experimental settings

Before modeling, the key-EIA was extracted from the original elementary image array (EIA) [44] according to the *interval*, which is the interval between two extracted EIs. For example, in Fig. 3, the marked EIs constitute the key-EIA with $interval = 5$. The key-EIA is then scanned into a PVS using the *Serpentine* form [45], as shown in Fig. 3(b). Similar to HEVC [46], the continuous frames, which are the EIs in the PVS, are processed based on groups, and the number of the frames in a certain group is the GOP value.

We choose the blocksize according to the balance between the bit consumption and the modeling effect. If the block is divided too small, it will not be efficient for coding. A certain EI in our experiments is $75 \times 75$, with the division shown in Fig. 3(b). If an EI is divided into $38 \times 38$ small blocks, we will obtain 4 blocks, and if the division is $19 \times 19$, there will be 16 blocks. In contrast to some transform-based coding methods using 4, 8, 16, and 32 blocks, the modeling-based methods can be implemented on blocks of any size. We do not use the traditional division because it will result in redundant parameters, which requires more bit consumption.

First, many Y-channel sequences were used for an exploratory experiment to determine the conditions where better effects could be achieved by both kernels. The quality and speed of modeling a $16 \times 20$ EIA sequence with different GOPs and CBs under the same number of models using EMR is displayed in Table I, from which it can be observed that the 75-based block has the worst performance and is the most time consuming. For the effect of GOP, based on the same blocksize, the experiments implemented under GOP = 8 and GOP = 16 consume more time and result in worse image quality than those under GOP = 4. Therefore, GOP = 8 and GOP = 16 are not comparable with GOP = 4 not only in terms of speed, but also in terms of visual effect. Overall, better conditions were selected for further experiments. For the Y channel, which is non-flat and filled with detailed pieces, the experimental conditions for both kernels were set as follows: GOP = 4 and CB = 19, 38, where the test numbers of models for each GOP = 4 sequence were 4, 8, 12, and 16 for CB = 19 and 16, 32, 48, and 64 for CB = 38. The divisions of the different CBs are shown in Fig. 3(b). Regarding the U/V channel, which was relatively flat, each EI was first downsampled into $38 \times 38$, and the corresponding experimental conditions were GOP = 4 and CB = 19, 38. Here, for CB = 19, the models were set at 2, 4, 6, and 8, and for CB = 38, the models were set at 16, 24, and 32. Therefore, it was possible to realize the same total number of models for the experimental EIA at four bitrates, which were equal for comparison.

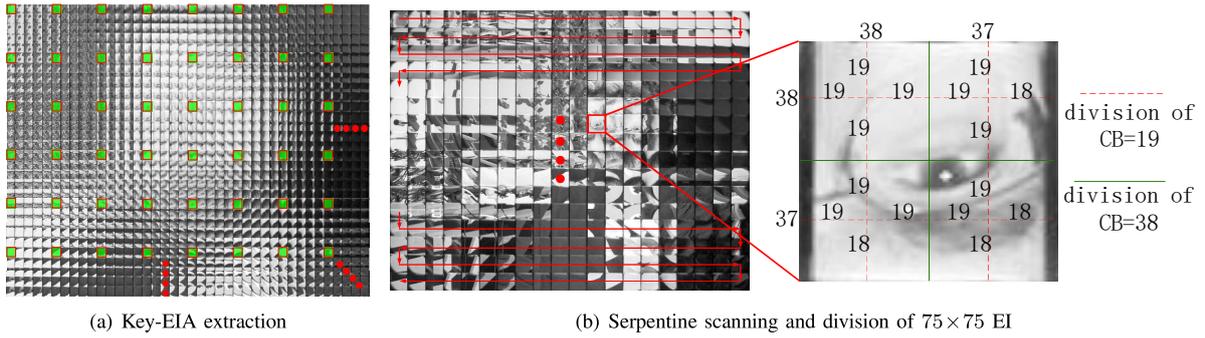

(a) Key-EIA extraction  (b) Serpentine scanning and division of $75 \times 75$ EI

Fig. 3. Example of *Fredo*. (a) Highlighted EIs are extracted into key-EIA with $interval = 5$. (b) Key-EIA scanning in serpentine order with the division of a certain EI.

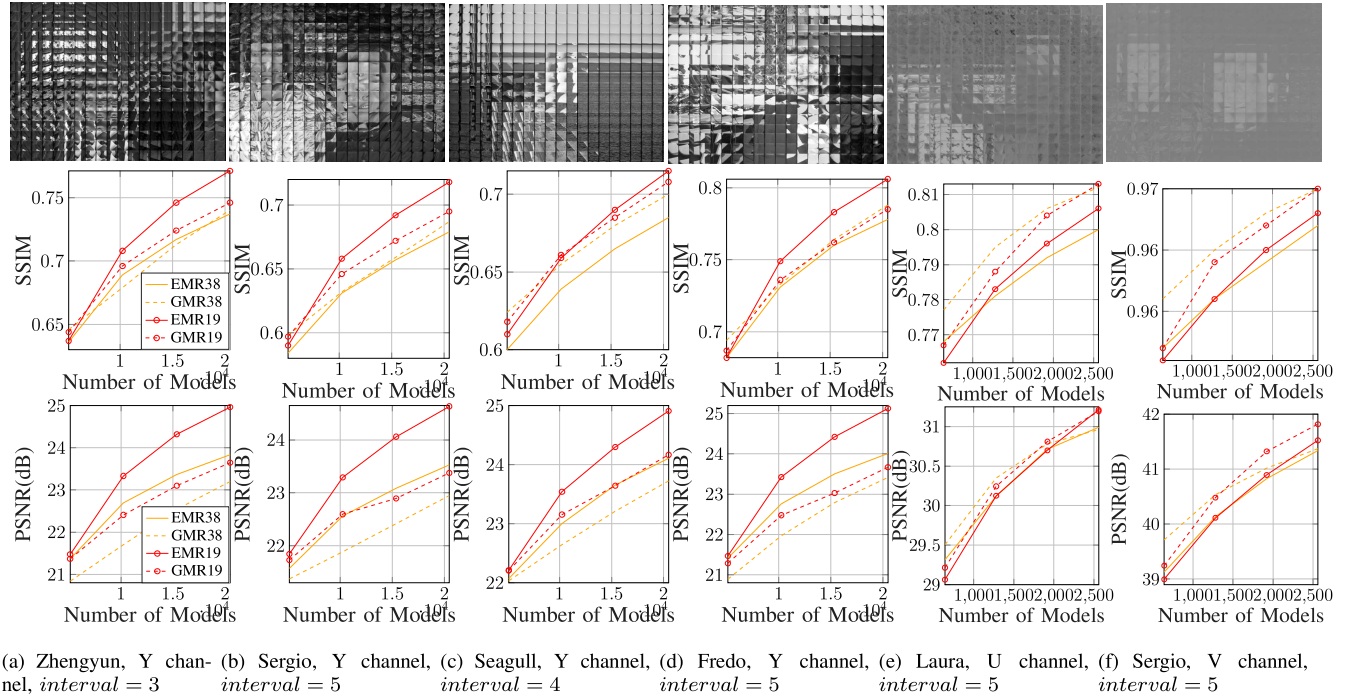

(a) Zhengyun, Y channel, $interval = 3$
(b) Sergio, Y channel, $interval = 5$
(c) Seagull, Y channel, $interval = 4$
(d) Fredo, Y channel, $interval = 5$
(e) Laura, U channel, $interval = 5$
(f) Sergio, V channel, $interval = 5$

Fig. 4. Experimental results of extracted-EI PVS. The representation of the legend is that EMR38 means EMR modeling for CB=38. EI-Y is $75 \times 75$ and EI-U/V is $38 \times 38$. The intervals of the given EIA are shown below.

### C. Result analysis

The results of the structural similarity index measure (SSIM) and peak signal-to-noise ratio (PSNR) are shown in Fig. 4, in which Fig. 4(a) to 4(d) show that EMR has tremendous advantages in non-adjacent EI sequences of the Y channel. Specifically, the CB = 19 based modeling has better behavior. Moreover, in Fig. 4(a) and 4(d), the performance of the EMR under CB = 38 can even be competitive with GMR under CB = 19. These advantages are distinct in the SSIM curves, and are more obvious in the PSNR comparisons. This result illustrates the advantage of using EMR for detailed image content. Regarding the U and V channels in Fig. 4(e) and 4(f), GMR is a little better than EMR under a smooth condition. However, combining the SSIM and PSNR results, CB = 38 is more efficient than CB = 19, which is different from the Y channel.

Above all, the advantages of EMR are more evident when the bitrates become higher, CB becomes smaller, and the texture varies more. Therefore, it can be preliminarily concluded that EMR can tolerate complex content changes and smaller correlations, while GMR prefers smoothness and continuity.

Therefore, in the following framework design, the Y channel is modeled using EMR under CB = 19 and GOP = 4, and the U/V channel is modeled using GMR under CB = 38 and GOP = 4.

## VI. Proposed Coding framework

A block diagram of the proposed framework is shown in Fig. 5. In the encoder, the inputs of the system are $I_0$, $\lambda$, and the initial $interval$. Parallax detection can determine the maximum $interval$ of the key-EIA, and 4-D AMLS can determine the best model combination for the key-EIA. In the decoder, the modeled key-EIA is restored, and the EIA $I_r$ output is reconstructed by merging and compensation. The four components in yellow boxes are jointly referred to as the *Linear Function Based Reconstruction* (LFBR) algorithm.

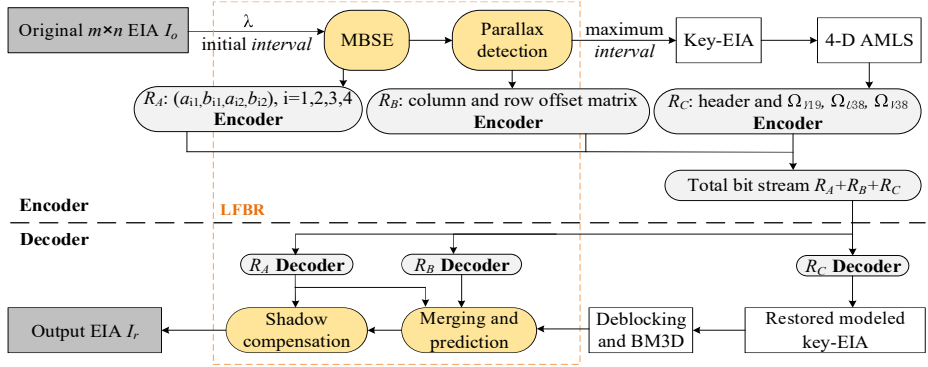

Fig. 5. Block diagram of our algorithm.

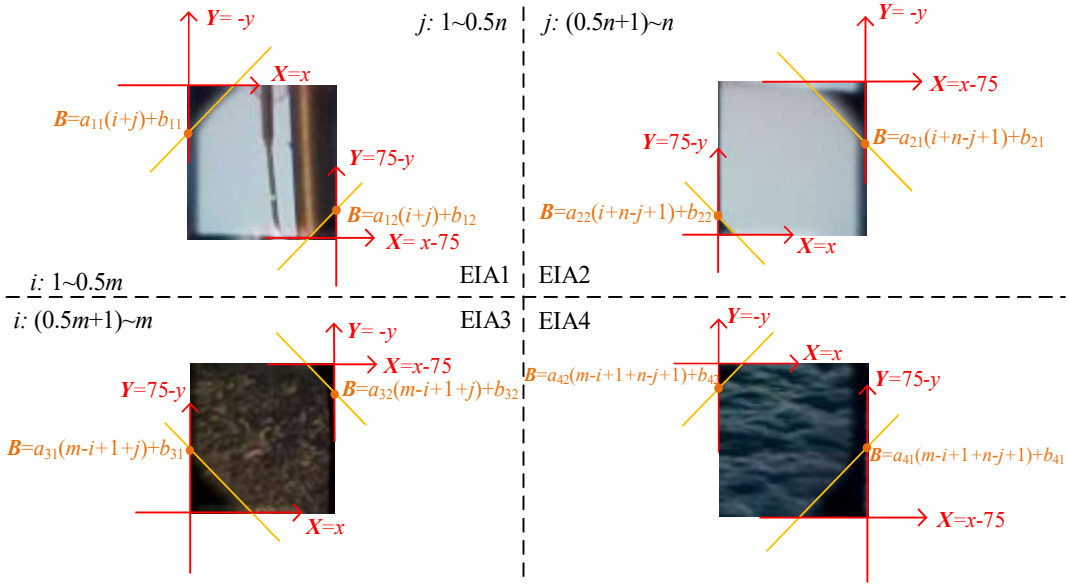

Fig. 6. Example of shadow extraction from *Seagull*. In (a), for EIA11, $a_{11}=0.64$, $b_{11}=-37$, $a_{12}=-0.5$, $b_{12}=18$; for EIA12, $a_{21}=0.64$, $b_{21}=-37$, $a_{22}=-0.35$, $b_{22}=18$; for EIA21, $a_{31}=-0.64$, $b_{31}=37$, $a_{32}=0.35$, $b_{32}=-18$; for EIA22, $a_{41}=-0.73$, $b_{41}=37$, $a_{42}=0.3$, $b_{42}=-18$.

### A. EIA Preprocessing

EIA preprocessing is necessary and contains two parts: the modeling-based shadow extraction (MBSE) algorithm and parallax detection. MBSE marks the shadow area of each EI. Therefore, in the reconstruction part, determining the shadow areas in EIs and completing the merger of adjacent EIs in the key-EIA are possible. Parallax detection marks the parallax between adjacent EIs, which determines the maximum interval key-EI extraction from the EIA to form the key-EIA.

*1) Modeling-based Shadow Extraction (MBSE) Algorithm:* This paper proposes an MBSE algorithm that can intelligently fit the border of a shadow area using a linear function, as shown in Fig. 6. The MBSE algorithm aims is to detect the shadow area from the original EIA in preparation for the decoder reconstruction. Coefficients $(a_{i1}, b_{i1}, a_{i2}, b_{i2})$, $i = 1, 2, 3, 4$ are lossless encoded as $R_A$ in Fig. 5.

First, the EIA is divided into four sections: EIA1, EIA2, EIA3, and EIA4. The numbers of rows and columns of a certain EI are defined as $i$ and $j$, respectively. The total numbers of rows and columns of the EIs for an EIA are $m$ and $n$, respectively. The border of a shadow corner is modeled as follows:

$$\boldsymbol{Y} = k\boldsymbol{X} + \boldsymbol{B}, \quad (34)$$

where $\boldsymbol{Y}$ is the coordinate variable of the pixel rows within a certain EI, and $\boldsymbol{X}$ is the coordinate variable of the pixel columns. $\boldsymbol{B}$ is the coordinate variable of the shadow intercept. $k = 1$ in EIA1 and EIA4, and $k = -1$ in EIA2 and EIA3.

Terms $x$ and $y$ are the column number and row number for a pixel of a certain EI, respectively. Then, the coordinate variables $\boldsymbol{Y}$ and $\boldsymbol{X}$ are actually the corresponding forms of $x$ and $y$ according to the EIA section of the EI, as shown in Fig. 6.

The representation of $\boldsymbol{B}$ from the shadow corner of several EIs is visualized in Fig. 7: A, B1, B2, and C show the three colors, and the overall comparison is displayed in C. Noticeably, when EI is closer to the boundary, the absolute value of $\boldsymbol{B}$ is larger. Therefore, $\boldsymbol{B}$ is assumed to be linearly dependent on the sum of the distances for an EI to the nearest boundary of the EIA in rows and columns. For example, the distance of B1 to the upper row is 1, and that to the right row is $n - (n-9) + 1 = 10$, whose sum is 11 and certainly $\boldsymbol{B} = 11a_{22} + b_{22}$.

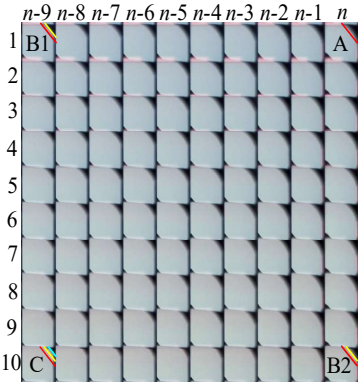

Fig. 7. Example of shadow extraction from *Seagull*.

After the modeling, the coefficients $(a_{i1}, b_{i1}, a_{i2}, b_{i2}), i = 1, 2, 3, 4$ can be tested for each specific EIA.

*2) Parallax Detection:* There are correlations between adjacent EIs of the EIA, from which it is possible to obtain offset matrices of rows and columns. These offset matrices can influence the maximum $interval$ as $interval \times max(offsets) <= 75$ and are necessary for reconstruction in the decoder. For encoding, the matrices are S-arranged, and their first-order differences are encoded as $R_B$ in Fig. 5 using *Adaptive Arithmetic Coding* (AAC) [47].

For the acquisition of the offset, while excluding the shadow area, the middle of an EI is considered, which is the $37 \times 45$ area marked in blue in Fig. 8. Next, a sliding window with a size of $37 \times 31$ is used to find the offset between adjacent rows or columns. Consider the red windows as an example: the left one is fixed at the position shown in Fig. 8 and the right one slides 15 pixels from the left to right within the blue box, with a different MSE at each step. The step corresponding to the minimal MSE is the desired offset. The step corresponding to the minimal MSE is the offset we want, and the yellow and green windows can visualize the offset matching.

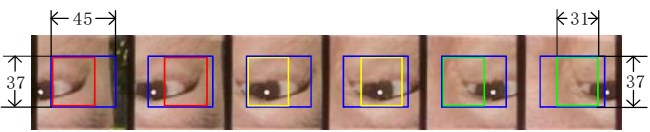

Fig. 8. Six adjacent EIs from *Sergio*.

Therefore, as for a $m \times n$ EIA, two offset matrices can be obtained, which are the $m \times (n-1)$ column offset matrix and $(m-1) \times n$ row offset matrix. Finally, the abrupt value of matrices must be adjusted and corrected according to the neighborhood.

### B. 4-D AMLS Modeling

The 4-D *Adaptive Model Selection* (4-D AMLS) is based on *Rate Distortion Optimization* (RDO) [48] and is used for determining the number of models for each PVS of the key-EIA, which is obtained the same manner as in Section V-B according to Fig. 3. According to Section V-A, a basic mode is selected for the key-EIA: EMR for Y channel with CB = 19 and GOP = 4 and GMR for downsampled U and V channels with CB = 38 and GOP = 4, which are all encoded as $R_c$ in Fig. 5.

Based on numerous experiments, the 14 parameters [$\mu_{X_j}$, $\mu_{Y_j}$, $\mu_{Z_j}$, $\mu_{W_j}$, $\Sigma_{X_j X_j}$, $\Sigma_{X_j Y_j}$, $\Sigma_{X_j Z_j}$, $\Sigma_{X_j W_j}$, $\Sigma_{Y_j Y_j}$, $\Sigma_{Y_j Z_j}$, $\Sigma_{Y_j W_j}$, $\Sigma_{Z_j Z_j}$, $\Sigma_{Z_j W_j}$, $\alpha_j$] can be quantized for each model with almost no loss as follows: [4,4,4,6,8,7,7,7,8,7,7,8,7,6] which is a total of 90 bits for Y (EMR19) and [4,4,4,6,5,4,4,4,5,4,4,5,5,4], which is a total of 62 bits for U and V (GMM38).

As for the Y channel, 4, 8, 12, and 16 EKs are used for each PVS with CB = 19 and GOP = 4. In addition, under GOP = 4, the downsampled $38 \times 38$ U and V channels are modeled using 2, 4, 8, and 16 Gaussian models. Then, R-D curves are obtained, as shown in Fig. 9, from which it is possible to estimate the range of the Lagrange parameter $\lambda = -\partial D / \partial R$ and obtain the minimum value of

$$J = D + \lambda R, \tag{35}$$

where D represents the sum of the squared error of all four frames in a PVS, and R represents the number of bits for modeling it. From Fig. 9, it can also be noticed that $\lambda$ of U/V is approximately half the size of Y. Therefore, we define $\lambda_Y = \lambda$ and $\lambda_{U/V} = 0.5\lambda$.

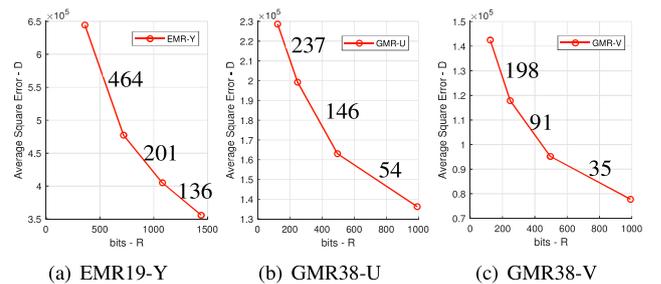

(a) EMR19-Y　　(b) GMR38-U　　(c) GMR38-V

Fig. 9. R-D curves of Y, U and V channel. $\lambda$s of the three segments of the curves are shown in the figures.

The pseudo-code of 4-D AMLS is presented in Algorithm 2, where matrix $\boldsymbol{J}$ represents the $J$-values of the modeled sequence with different numbers of models, and $cell(\boldsymbol{\Omega})$ stores the parameters of all the model conditions. For example, $J_{E_{15}}$ in Step 3 denotes the $J$-value of a 19-based 4-frames PVS modeled by the EK using 15 models, whose parameters are recorded in $\Omega_{E_{15}}$. The optimal parameter set is obtained according to the minimum value of $\boldsymbol{J}$.

### C. Coding of Modeling Parameters

After 4-D AMLS, the modeling parameters are encoded. For a certain channel of an EIA, a total header should be added to mark span $E$ and minimum value $M$ of each parameter. Specifically, 28 marks for Y19 and 20 marks for U/V38 should be added. All of the marks are losslessly encoded using arithmetic coding. Each type of parameter is encoded according to Table II using AAC after the quantization. The kernel bit allocation is explained in the following subsections.

*1) Cholesky decomposition of the position matrix:* As for the $3 \times 3$ semi-positive covariance matrix $\boldsymbol{R}_j$, Cholesky

TABLE II
KERNEL BITS ALLOCATION

| | NM | $n_{NM}$ | Bits Allocation (bits) | | | | | | | | | | | | | |
|---|---|---|---|---|---|---|---|---|---|---|---|---|---|---|---|---|
| | | | Parameter Bits - $n_{PB}$ | | | | | | | | | | | | | |
| | | | $n_{\mu_{X_j}}$ | $n_{\mu_{Y_j}}$ | $n_{\mu_{Z_j}}$ | $n_{\mu_{W_j}}$ | $n_{u_{11_j}}$ | $n_{u_{12_j}}$ | $n_{u_{13_j}}$ | $n_{u_{22_j}}$ | $n_{u_{23_j}}$ | $n_{u_{33_j}}$ | $n_{\Sigma_{X_j W_j}}$ | $n_{\Sigma_{Y_j W_j}}$ | $n_{\Sigma_{Z_j W_j}}$ | $\alpha_j$ |
| Y19 | >1 | 4 | 4 | 4 | 4 / 5 | 6 | 7 | 6 | 6 | 7 | 6 | 7 | 6 | 6 | 6 | 5 |
| | =1 | 4 | - | - | - | 6 | - | - | - | - | - | - | 6 | 6 | 6 | - |
| U/V38 | >1 | 3 | 4 | 4 | 4 / 5 | 5 | 4 | 3 | 3 | 4 | 3 | 4 | - | - | - | - |
| | =1 | 3 | - | - | - | 5 | - | - | - | - | - | - | - | - | - | - |

$^a$ $n_{\mu_{Z_j}}$ is 4 when $\lambda \geq 300$, and 5 when $\lambda < 300$. The $2^{n_{\mu_{Z_j}}}$ quantization points of $\mu_{Z_j}$ include the integer points 1, 2, 3, and 4 (GOP = 4).
$^b$ $NM$ is the number of models and $n_{NM}$ represents its bit consumption. The term $n_{PB}$ shows the bits allocation for parameters of a model.
$^c$ Regarding quantization of the U/V channel, $\Sigma_{X_j W_j}$, $\Sigma_{Y_j W_j}$, and $\Sigma_{Z_j W_j}$ are constantly 0. In addition, $\alpha_j$ is set to be $1/NM$ for the U and V channels.

---

**Algorithm 2** 4-D AMLS Algorithm
---
1: **Input:** Y: 19×19×4 (or 19×18×4, 18×19×4, 18×18×4) PVS of EI or
　　　　U/V: 38×38×4 PVS of EI
　　　　and a $\lambda$
2: **if** Y
3: 　 Calculate $J$ in (35) of sequence Y at 1 to 16 models and find the parameters corresponding to the minimum $J$. $\lambda_Y = \lambda$, $\boldsymbol{J} = \begin{bmatrix} J_{E_1} J_{E_2} \cdots J_{E_{16}} \end{bmatrix}$ and $cell(\boldsymbol{\Omega}) = \begin{bmatrix} \boldsymbol{\Omega}_{E_1} \boldsymbol{\Omega}_{E_2} \cdots \boldsymbol{\Omega}_{E_{16}} \end{bmatrix}$. $\boldsymbol{\Omega}_{Y19} = \boldsymbol{\Omega}\{\text{find }(\boldsymbol{J} == \min \boldsymbol{J})\}$.
4: **elseif** U or V
5: 　 Similarly, we can obtain $\lambda_{U/V} = 0.5\lambda$, $\boldsymbol{J} = \begin{bmatrix} J_{G_1} J_{G_2} \cdots J_{G_8} \end{bmatrix}$ and $cell(\boldsymbol{\Omega}) = \begin{bmatrix} \boldsymbol{\Omega}_{G_1} \boldsymbol{\Omega}_{G_2} \cdots \boldsymbol{\Omega}_{G_8} \end{bmatrix}$. $\boldsymbol{\Omega}_{U/V38} = \boldsymbol{\Omega}\{\text{find }(\boldsymbol{J} == \min \boldsymbol{J})\}$.
6: **end**
7: **Output:** Optimal parameters $\boldsymbol{\Omega}_{Y19}$ or $\boldsymbol{\Omega}_{U/V38}$.
---

decomposition can be conducted as follows: $\boldsymbol{R}_j = \boldsymbol{U}^T \boldsymbol{U}$, where $\boldsymbol{U}$ is an upper triangular matrix:

$$\boldsymbol{U} = \begin{bmatrix} u_{11} & u_{12} & u_{13} \\ 0 & u_{22} & u_{23} \\ 0 & 0 & u_{33} \end{bmatrix}. \quad (36)$$

Therefore, only the six parameters in (36) are encoded, through which it is possible to narrow the range of parameters, as well as keep $\boldsymbol{R}_j$ semi-positive.

*2) Special case of the single-kernel coding:* If a block is optimally represented by one model, the mean value $\hat{\boldsymbol{\mu}}_1$ and covariance matrix $\boldsymbol{R}_1$ are only determined by the size of the block sequence and do not need to be encoded. This is because of the particularity of the EM algorithm in Algorithm 1: after parameter initialization, we then reach the result of posterior probability $Qt_{ij}$ (E-step) utilizing Equation (33). When the number of models $j = 1$, $Qt_{i1} = 1$, the M-step turns out to be

$$\boldsymbol{\mu}_1 = \frac{\sum_{i=1}^{N} \boldsymbol{\varphi}_i}{N}, \quad \boldsymbol{\Sigma}_1 = \frac{\sum_{i=1}^{N}(\boldsymbol{\varphi}_i - \boldsymbol{\mu}_1)'(\boldsymbol{\varphi}_i - \boldsymbol{\mu}_1)}{N-1}, \quad \alpha_1 = 1. \quad (37)$$

In this case, the iteration results are directly dependent on the original samples $\boldsymbol{\varphi}_i$. For a certain sequence size, the 3-D position samples are always the same, and only the gray values are differently distributed. Therefore, the position mean vector $\hat{\boldsymbol{\mu}}_1$ and the position covariance matrix $\boldsymbol{R}_1$ are determined.

Only $\mu_{W_1}$, $\Sigma_{X_1 W_1}$, $\Sigma_{Y_1 W_1}$, and $\Sigma_{Z_1 W_1}$ are encoded according to Table II.

*3) Parameter Decoding:* In the decoder, with $\mu_{X_j}$ as a quantization example, $n_{\mu_{X_j}}$ is the number of bits of $\mu_{X_j}$ known from Table II, $k$ is the encoded bit representation, and the $k$th restored parameters $\mu_{X_j}$ is calculated as follows:

$$\mu_{X_j}[k] = M_{\mu_X} + (k-1) \cdot \frac{E_{\mu_X}}{2^{n_{\mu_{X_j}}} - 1}, k = 1, 2, \cdots, 2^{n_{\mu_{X_j}}}. \quad (38)$$

The decoded parameter sets of all the GOP = 4 sequences are used to reconstruct the modeling key-EIA using (2). After decoding, de-blocking is a necessary postprocessing step. All the block borders are smoothed by copying the boundary pixel and overlapping in proportion. For all channels, the duplicate pixel width of each key-EIA is 3. Then, a *Block-matching and 3D filtering* (BM3D) algorithm [49] is finally added, whose $\sigma = 15$ for Y and $\sigma = 20$ for U and V. Finally, the U and V channels are upsampled to the same size as Y.

*D. Reconstruction*

*1) Merging and Prediction:* First, the reconstruction starts from the adjacent EIs of the key-EIA in the columns followed by the rows. An example of the reconstruction process is shown in Fig. 10.

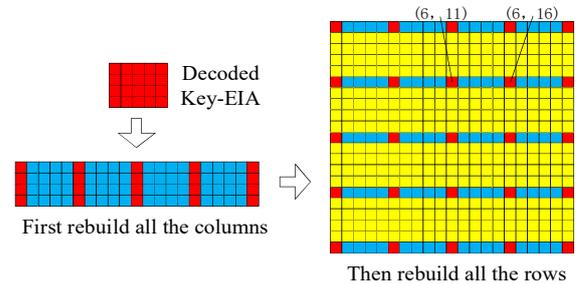

Fig. 10. Example of the reconstruction process. The red EIA represents the decoded key-EIA. The blue EIs represent the ones that are reconstructed by key-EIA in columns. Finally, the other rows (yellow ones) are predicted through the reconstructed rows (red and blue ones).

Two adjacent EIs are selected from the decoded key-EIA of *Fredo*, whose position $(i,j)$ in the reconstructed EIA should be (6,11) and (6,16), which are marked in Fig. 10. Then, the reconstruction example for the EIs between (6,11) and (6,16) is shown in Fig. 11, whose parameters have been encoded as $R_A$.

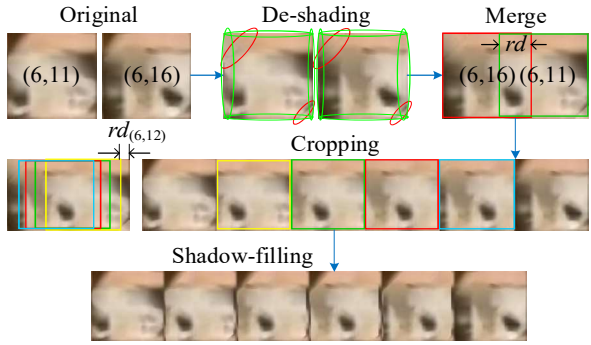

Fig. 11. Reconstruction example of EI (6,12) to EI (6,15) from decoded key-EIA of *Fredo*. The reconstruction is implemented the same steps under Y, U, and V.

As shown in Fig. 11, the six columns of pixels on the left and right sides (or four rows of pixels on the top and bottom sides) of the EIs in the key-EIA should be compensated, as shown in the green circles of the de-shading step in Fig. 11. Meanwhile, the shadow corners circled in red are compensated with location $(i, j)$ according to the linear function in Fig. 6. The compensation should be referred to the gray value 3-pixels outside the shadow edge toward the center of the EI.

$rd_{(i,j)}$ is the offset of EI(i,j) and its former column (row). Then, $rd = 75 - rd_{(6,12)} - rd_{(6,13)} - rd_{(6,14)} - rd_{(6,15)} - rd_{(6,16)}$ in Fig. 11, where $rd_{(6,12)}$ to $rd_{(6,16)}$ are obtained from the decoded offset matrix encoded as $R_B$. As for merging, the overlapping area is obtained through gray values according to their proportion. For each row of pixels, the left shadow area for (6,11) is weighted as 0, and the right shadow area for (6,16) is weighted as 0, whereas the others are scaled by $rd$. Further, the cropping step is used to remove the unknown EIs from the merging image according to each offset.

*2) Shadow Compensation:* Finally, the corner shadow is reconstructed by copying the area from (6,11), whose size is determined by the linear function of the shadow of the corresponding EI. In addition, the seams between rows and columns are also restored by referring to the seams of the key-EIA.

## VII. EXPERIMENTAL RESULTS

All the test images had a size of $7200 \times 5400$ in the experiments. The experimental equipment has an Intel(R) Core(TM) i7-7700K CPU @ 4.20GHz. The proposed algorithm was implemented under MATLAB R2017b, and different bits per pixel (bpps) were realized by adjusting $\lambda$ and *interval* as follows: $\lambda = 1000$, $interval = 5$; $\lambda = 300$, $interval = 5$; $\lambda = 150$, $interval = 4$; and $\lambda = 75$, $interval = 3$. To implement the HEVC, the EIA was rearranged into an EI pseudo sequence and encoded with GOP = 4 and GOP = 16 under a low delay P frame. The statistics were selected from QP = 51, 50, 49, 47, 45, 43, and 41. Additionally, for the SAI-based pseudo sequence, HEVC GOP = 4 was implemented under QP = 51, 43, and 40; and HEVC GOP = 16 was implemented under QP = 51, 40, and 38. All the experiments of HEVC were implemented using the Visual Studio 2013 platform. The input of JPEG 2000 is an entire image, either on the EI array or the SAI array. JPEG 2000 was implemented under 0.01 bpp, 0.02 bpp, 0.03 bpp, 0.04 bpp, 0.05 bpp, and 0.06 bpp using MATLAB R2017b. The calculations of PSNR and SSIM for the reconstructed three-channel images for all the algorithms are

$$PSNR = 10 \log_{10}(\frac{255^2}{\frac{1}{3}(MSE_Y + MSE_U + MSE_V)}), \quad (39)$$

$$SSIM = \frac{1}{3}(SSIM_Y + SSIM_U + SSIM_V), \quad (40)$$

where $MSE_Y$, $MSE_U$, and $MSE_V$ are the MSE of the Y, U, and V channels, respectively. $SSIM_Y$, $SSIM_U$, and $SSIM_V$ are the SSIM of the Y, U, and V channels, respectively.

As shown in Fig. 12, at low bitrates, the proposed method can achieve the same visual effect over *HEVC Low Delay P* for GOP = 4 and GOP = 16 with only 20%–30% of the bits consumed, and the compression efficiency is consistently ahead of JPEG 2000 at less than 0.05 bpp. Additionally, the SSIM is generally around 0.85, which reflects the acceptable visual effect of the proposed method. Fig. 14 represents the SSIM comparison of the rendered views, which are obtained by extracting the central $8 \times 8$ blocks of all the EIs and stitching them together [5]. The advantages of the proposed method in Fig. 14 is more obvious than that in Fig. 12. Therefore, the proposed method can realize a favorable effect in the rendered views. However, this method is not that superior according to the PSNR results, as shown in Fig. 13. For both the SSIM and PSNR, the SAI-based coding results have poor behaviors.

Rendered views of two images are shown in Fig. 15 and Fig. 16. The proposed modeling-based compression shows clear outlines and a gentle color, whereas the HEVC results contain significant noise but more high-frequency components. As for JPEG 2000, it shows greater details than the proposed method, but is also a little unclear. However, the SAI-based results are the worst among others. Overall, the proposed algorithm is found to be an applicable method under low bit rates. Moreover, from the results, the images reconstructed using the proposed algorithm are clear and color adaptive, making them suitable for human vision. However, the proposed algorithm has poor behavior at high bitrates because the model can represent a set of pixels without specific high-frequency information. Conversely, the traditional transform-based algorithm can preserve the details well. Therefore, the modeling-based algorithm is poor at representing detailed image content, which makes its behavior at high bitrates inferior to the traditional methods. This is also the reason why the PSNRs of the proposed method are not competitive.

A time consumption comparison in Table III shows that the proposed method consumed more time than HEVC and JPEG 2000 required the least amount of time.

### TABLE III
Time consumption at several bpps of different algorithms

|  | Fredo | Jeff | Seagull | Sergio | Zhengyun | Laura |
|---|---|---|---|---|---|---|
| bits per pixel (bpp) | 0.053 | 0.054 | 0.051 | 0.060 | 0.05 | 0.07 |
| Ours | 10968s | 11114s | 11122s | 11447s | 11050s | 12139s |
| HEVC GOP = 4-EI | 5068s | 5703s | 4610s | 5652s | 5222s | 6139s |
| JPEG 2000-EI | 17.67s | 17.40s | 18.41s | 19.35s | 17.84s | 19.46s |

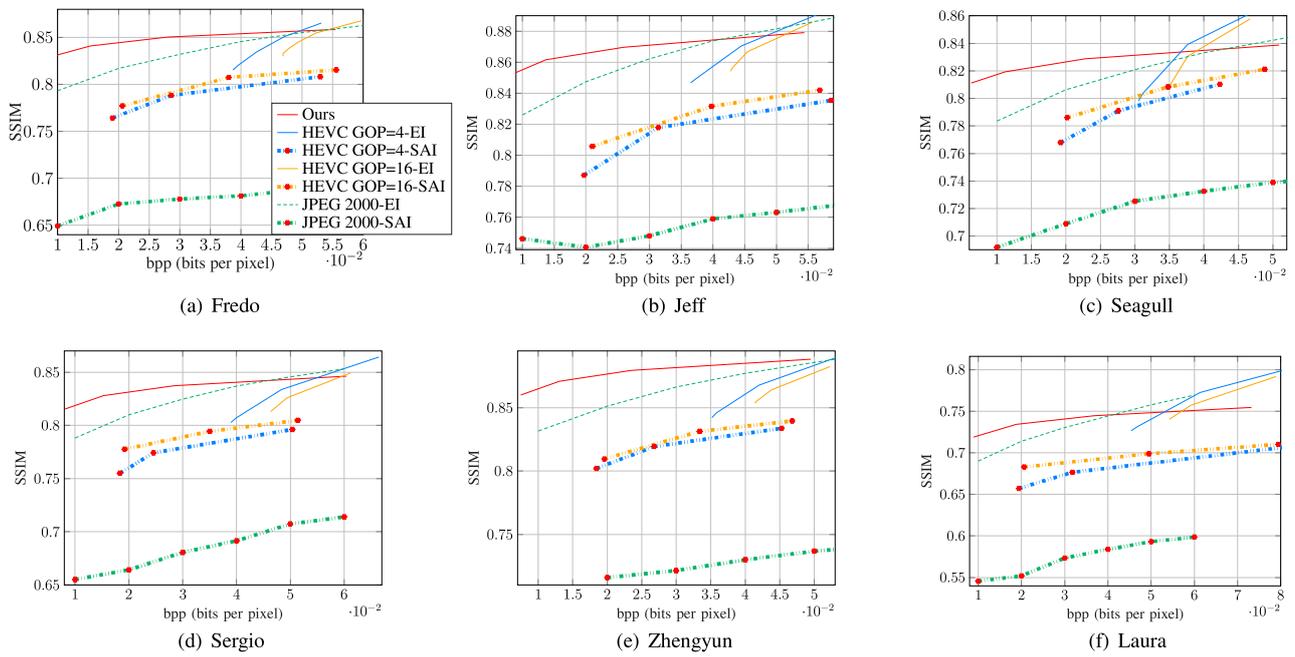

Fig. 12. The bpp-SSIM maps of EIA towards different algorithms. The EI-based and SAI-based coding for HEVC and JPEG 2000 are all given.

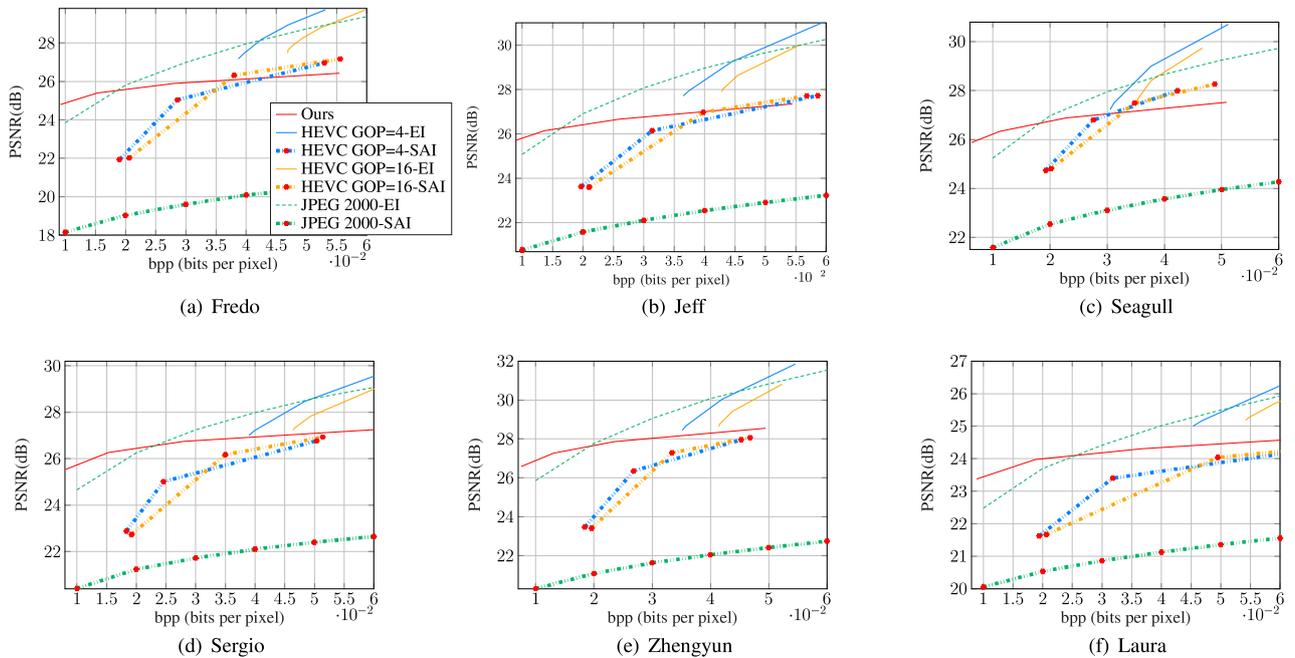

Fig. 13. The bpp-PSNR(dB) maps of EIA towards different algorithms. The EI-based and SAI-based coding for HEVC and JPEG 2000 are all given.

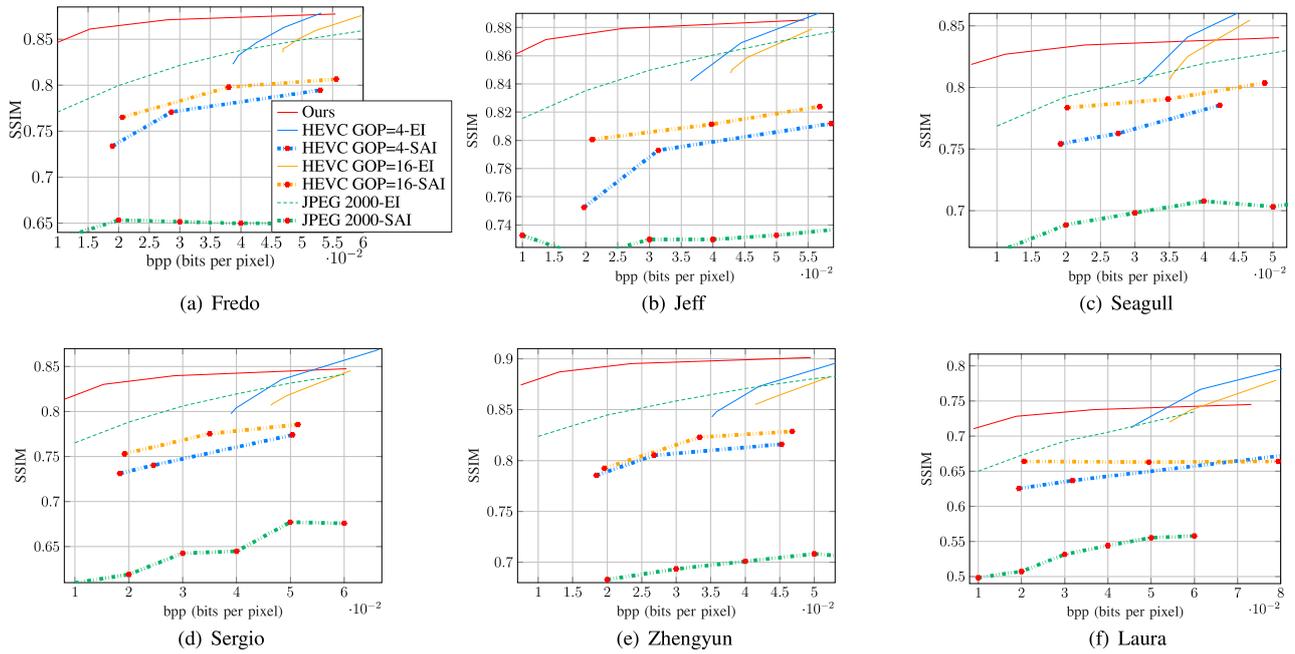

Fig. 14. The bpp-SSIM maps of the central rendered view (composed of the central $8 \times 8$ block of each EI) towards different algorithms. The EI-based and SAI-based coding for HEVC and JPEG 2000 are all given.

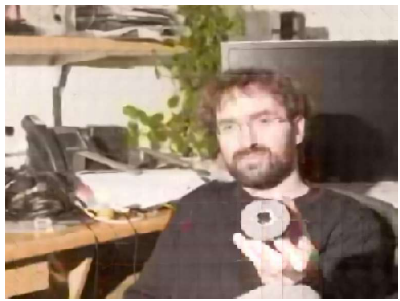
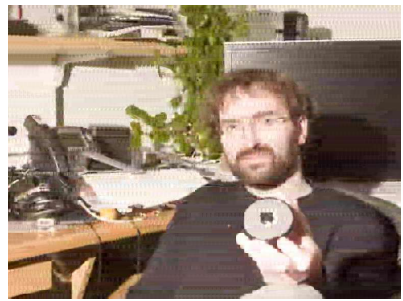
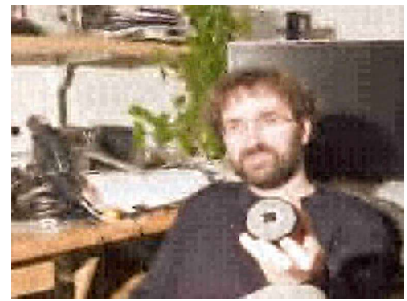

(a) Ours 0.015bpp/28.37dB/0.86  (b) HEVC$_{GOP=4}$ (EI Seq.) 0.038bpp/28.88dB/0.82  (c) JPEG 2000 (EI Seq.) 0.02bpp/26.98dB/0.80

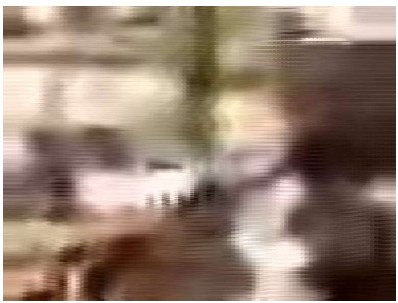
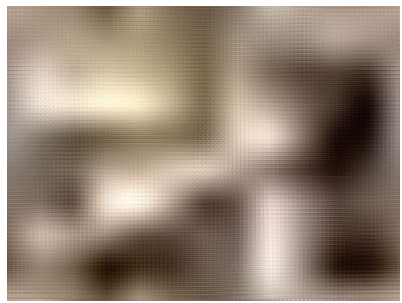

(d) HEVC$_{GOP=4}$ (SAI Seq.) 0.019bpp/21.63dB/0.73  (e) JPEG 2000 (SAI Seq.) 0.02bpp/18.77dB/0.65

Fig. 15. The rendered views of the decoded results of different algorithms for *Fredo*. The PSNR/SSIMs of the given rendered views are shown below.

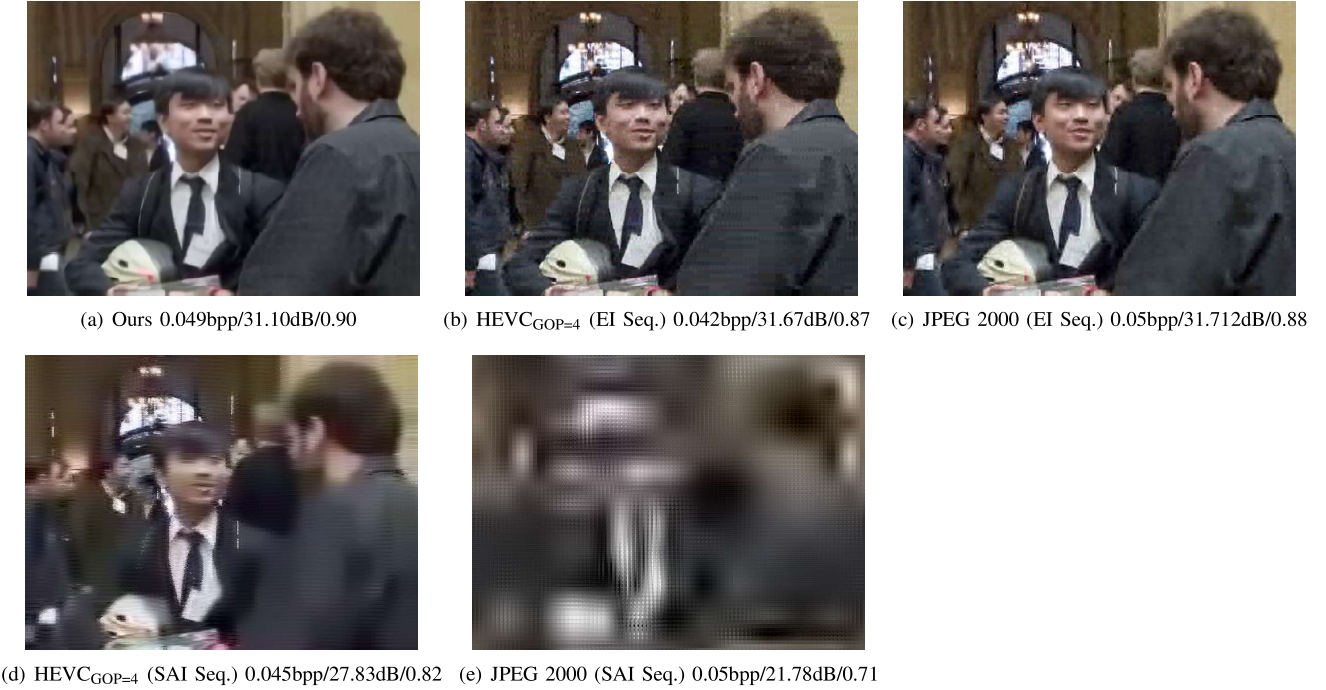

(a) Ours 0.049bpp/31.10dB/0.90    (b) HEVC$_{\text{GOP}=4}$ (EI Seq.) 0.042bpp/31.67dB/0.87    (c) JPEG 2000 (EI Seq.) 0.05bpp/31.712dB/0.88

(d) HEVC$_{\text{GOP}=4}$ (SAI Seq.) 0.045bpp/27.83dB/0.82    (e) JPEG 2000 (SAI Seq.) 0.05bpp/21.78dB/0.71

Fig. 16. The rendered views of the decoded results of different algorithms for *Zhengyun*. The PSNR/SSIMs of the given rendered views are shown below.

### A. The comparison with related work [5]

The HEVC framework improved with GMR prediction was proposed in [5], which has huge advantages over HEVC and it is even competitive with HEVC-Screen Content Coding (HEVC-SCC), an advanced version of HEVC. The coding of two $6 \times 11$ EIA of *Seagull* and *Zhengyun* are provided for examples. For comparison, the QPs of [5] are 51, 49, 47, 45, 41, and 35 for comparison. The results are shown in Fig. 17, from which we can observe that the coding effect of our algorithm is far from that of [5].

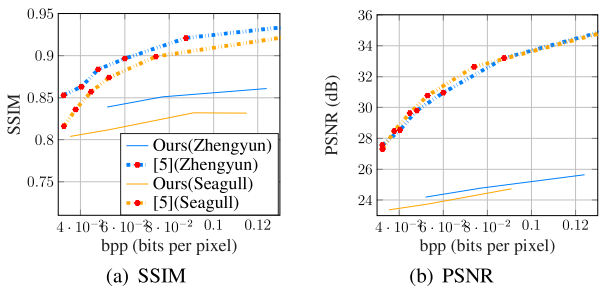

(a) SSIM    (b) PSNR

Fig. 17. The comparison results of the coding of two $6 \times 11$ EIA between our algorithm and [5].

## VIII. CONCLUSION AND FUTURE WORK

This paper proposed a completely modeling-based framework for LF image modeling, which is a breakthrough for LF image compression. Moreover, the modeling theory based on the 4-D EK is brand new.

According to the experiment results, the proposed 4-D EK and its correlated theory behaved well in the PVS modeling of LF images, especially in the most essential Y channel. The proposed 4-D EMR performed very well for EIA compression. In addition, the LFBR reconstruction algorithm was also an excellent prediction framework for LF images. Moreover, the coding efficiency was competitive with that of the standard algorithms (HEVC and JPEG 2000) in terms of the SSIM. Although the PSNR behavior was not superior to that of the HEVC, compared to transform-based algorithms, the visual effect was acceptable without much noise. In essence, the framework was a completely modeling-based algorithm and enriched the practical significance of using a 4-D EK in PVS coding, which is a breakthrough for Gaussian modeling-based compression.

The derivation process reported in this paper could be used as a reference for other kernel functions and can be applied to regression modeling in the future. However, the modeling approach showed poor behavior at high bitrates, which can be improved in the future. For example, better parameter optimization for EMR may improve the modeling results, and a suitable combination of residual signals for modeling-based coding can also compensate the deficiency in high bitrates. Furthermore, we are anticipating additional applications in other fields based on Epanechnikov-correlated theories.

## APPENDIX A
### CALCULATION OF THE FOUR INTEGRALS

In order to solve the integral $\iiiint_{\Omega_{\hat{\varphi}}} \left(1 - \hat{\varphi}^T \mathbf{\Lambda}^2 \hat{\varphi}\right) d\Omega_{\hat{\varphi}}$, we need to convert the variable of standard coordinates $(\hat{x}, \hat{y}, \hat{z}, \hat{w})$ to that of spherical coordinates $(r, \alpha, \beta, \theta)$

$$\begin{cases} \hat{x} = ar \sin \alpha \sin \beta \cos \theta \\ \hat{y} = br \sin \alpha \sin \beta \sin \theta \\ \hat{z} = cr \sin \alpha \cos \beta \\ \hat{w} = dr \cos \alpha \end{cases} \quad (41)$$

in which $0 \leq r \leq 1, 0 \leq \alpha \leq \pi, 0 \leq \beta \leq \pi, 0 \leq \theta \leq 2\pi$, and

$$J = \frac{\partial(\hat{x}, \hat{y}, \hat{z}, \hat{w})}{\partial(r, \alpha, \beta, \theta)} = \begin{vmatrix} \frac{\partial \hat{x}}{\partial r} & \frac{\partial \hat{x}}{\partial \alpha} & \frac{\partial \hat{x}}{\partial \beta} & \frac{\partial \hat{x}}{\partial \theta} \\ \frac{\partial \hat{y}}{\partial r} & \frac{\partial \hat{y}}{\partial \alpha} & \frac{\partial \hat{y}}{\partial \beta} & \frac{\partial \hat{y}}{\partial \theta} \\ \frac{\partial \hat{z}}{\partial r} & \frac{\partial \hat{z}}{\partial \alpha} & \frac{\partial \hat{z}}{\partial \beta} & \frac{\partial \hat{z}}{\partial \theta} \\ \frac{\partial \hat{w}}{\partial r} & \frac{\partial \hat{w}}{\partial \alpha} & \frac{\partial \hat{w}}{\partial \beta} & \frac{\partial \hat{w}}{\partial \theta} \end{vmatrix} = abcd r^3 \sin^2 \alpha \sin \beta. \quad (42)$$

Thus, the integral turns out to be

$$\begin{aligned}
&\iiiint\limits_{\Omega_{\hat{\varphi}}: \hat{\varphi}^T \Lambda \hat{\varphi} \leq 1} \left(1 - \hat{\varphi}^T \Lambda \hat{\varphi}\right) d\Omega_{\hat{\varphi}} \\
&= \iiiint\limits_{\frac{\hat{x}^2}{a^2} + \frac{\hat{y}^2}{b^2} + \frac{\hat{z}^2}{c^2} + \frac{\hat{w}^2}{d^2} \leq 1} \left[1 - \left(\frac{\hat{x}^2}{a^2} + \frac{\hat{y}^2}{b^2} + \frac{\hat{z}^2}{c^2} + \frac{\hat{w}^2}{d^2}\right)\right] d\hat{x} d\hat{y} d\hat{z} d\hat{w} \\
&= abcd \int_0^{2\pi} d\theta \int_0^{\pi} d\alpha \int_0^{\pi} d\beta \int_0^1 \left(1 - r^2\right) r^3 \sin^2 \alpha \sin \beta dr \\
&= \frac{\pi^2 abcd}{6} = \frac{\pi^2}{6|\Lambda|}.
\end{aligned} \quad (43)$$

## APPENDIX B
## CALCULATION OF $\hat{\Sigma}$

$\hat{X}, \hat{Y}, \hat{Z}$, and $\hat{W}$ are the random value of $\hat{x}, \hat{y}, \hat{z}$, and $\hat{w}$, therefore the covariance matrix of $\hat{f}(\hat{\varphi})$ is

$$\hat{\Sigma} = \begin{bmatrix} \text{Cov}(\hat{X},\hat{X}) & \text{Cov}(\hat{X},\hat{Y}) & \text{Cov}(\hat{X},\hat{Z}) & \text{Cov}(\hat{X},\hat{W}) \\ \text{Cov}(\hat{Y},\hat{X}) & \text{Cov}(\hat{Y},\hat{Y}) & \text{Cov}(\hat{Y},\hat{Z}) & \text{Cov}(\hat{Y},\hat{W}) \\ \text{Cov}(\hat{Z},\hat{X}) & \text{Cov}(\hat{Z},\hat{Y}) & \text{Cov}(\hat{Z},\hat{Z}) & \text{Cov}(\hat{Z},\hat{W}) \\ \text{Cov}(\hat{W},\hat{X}) & \text{Cov}(\hat{W},\hat{Y}) & \text{Cov}(\hat{W},\hat{Z}) & \text{Cov}(\hat{W},\hat{W}) \end{bmatrix}. \quad (44)$$

We take the calculation of $\text{Cov}(X, X)$ as an example:

$$\text{Cov}\left(\hat{X}, \hat{X}\right) = E\left(\hat{X}^2\right) - E^2\left(\hat{X}\right). \quad (45)$$

$$E(\hat{X}) = \iiiint\limits_{\hat{\varphi}^T \Lambda^{-1} \hat{\varphi} \leq 1} \hat{x} \frac{6|\Lambda|(1 - \hat{\varphi}^T \Lambda^{-1} \hat{\varphi})}{\pi^2} d\hat{\varphi} = 0 \quad (46)$$

$$\begin{aligned}
E(\hat{X}^2) &= \iiiint\limits_{\hat{\varphi}^T \Lambda^{-1} \hat{\varphi} \leq 1} \hat{x}^2 \frac{6|\Lambda|(1 - \hat{\varphi}^T \Lambda^{-1} \hat{\varphi})}{\pi^2} d\hat{\varphi} \\
&= \frac{6}{\pi^2} \int_0^1 dr \int_0^{2\pi} d\theta \int_0^{\pi} d\alpha \int_0^{\pi} (ar \sin \alpha \sin \beta \cos \theta)^2 \\
&\qquad (1 - r^2) r^3 \sin^2 \alpha \sin \beta d\beta \\
&= \frac{a^2}{8},
\end{aligned} \quad (47)$$

Thus, $\text{Cov}(\hat{X}, \hat{X}) = a^2/8$. Similarly, we can know the others as $\text{Cov}(\hat{Y}, \hat{Y}) = b^2/8$, $\text{Cov}(\hat{Z}, \hat{Z}) = c^2/8$, $\text{Cov}(\hat{W}, \hat{W}) = d^2/8$, and other coefficients in $\hat{\Sigma}$ are all 0. Therefore,

$$\hat{\Sigma} = \text{diag}\left(\frac{a^2}{8}, \frac{b^2}{8}, \frac{c^2}{8}, \frac{d^2}{8}\right) = \frac{1}{8}\left(\Lambda^{-1}\right)^2. \quad (48)$$

## ACKNOWLEDGMENT

The authors would like to thank Deyang Liu for his help to this paper.

## REFERENCES

[1] Wu, G., Masia, B., Jarabo, A., Zhang, Y., Wang, L., and Dai, Q., et al. "Light field image processing: an overview". *IEEE Journal of Selected Topics in Signal Processing*, 2017, 11(7), 926-954.

[2] Aggoun A, Tsekleves E, Swash M R, et al. "Immersive 3D Holoscopic Video System"[J]. *IEEE MultiMedia*, 2013, 20(1):28-37.

[3] M. Magnor and B. Girod, "Data compression for light-field rendering," in *IEEE Transactions on Circuits and Systems for Video Technology*, vol. 10, no. 3, pp. 338-343, April 2000, doi: 10.1109/76.836278.

[4] Liu, Deyang, et al. "3D holoscopic image coding scheme using HEVC with Gaussian process regression." *Signal Processing Image Communication* 47(2016):438-451.

[5] D. Liu, P. An, R. Ma, W. Zhan, X. Huang and A. A. Yahya, "Content-Based Light Field Image Compression Method With Gaussian Process Regression," in *IEEE Transactions on Multimedia*, vol. 22, no. 4, pp. 846-859, April 2020, doi: 10.1109/TMM.2019.2934426.

[6] Y. Li, M. Sjöström, R. Olsson and U. Jennehag, "Coding of Focused Plenoptic Contents by Displacement Intra Prediction," in *IEEE Transactions on Circuits and Systems for Video Technology*, vol. 26, no. 7, pp. 1308-1319, July 2016, doi: 10.1109/TCSVT.2015.2450333.

[7] L. Li, Z. Li, B. Li, D. Liu and H. Li, "Quadtree-Based Coding Framework for High-Density Camera Array-Based Light Field Image," in *IEEE Transactions on Circuits and Systems for Video Technology*, vol. 30, no. 8, pp. 2694-2708, Aug. 2020, doi: 10.1109/TCSVT.2019.2924313.

[8] J. Hou, J. Chen and L. Chau, "Light Field Image Compression Based on Bi-Level View Compensation With Rate-Distortion Optimization," in *IEEE Transactions on Circuits and Systems for Video Technology*, vol. 29, no. 2, pp. 517-530, Feb. 2019, doi: 10.1109/TCSVT.2018.2802943.

[9] Verhack. Ruben, et al. "Steered mixture-of-experts for light field coding, depth estimation, and processing", *2017 IEEE International Conference on Multimedia and Expo (ICME) IEEE*, 2017.

[10] Verhack. Ruben, et al. "Steered Mixture-of-Experts for Light Field Images and Video: Representation and Coding", *IEEE Transactions on Multimedia* PP.99(2019):1-1.

[11] Verhack. Ruben, et al. "A universal image coding approach using sparse steered Mixture-of-Experts regression." 2016 *IEEE International Conference on Image Processing (ICIP)* IEEE, 2016.

[12] L. Lange, R. Verhack and T. Sikora, "Video representation and coding using a sparse steered mixture-of-experts network," *2016 Picture Coding Symposium (PCS)*, 2016, pp. 1-5, doi: 10.1109/PCS.2016.7906369.

[13] Mesquita, D. P. P., J. P. P. Gomes, and A. H. S. Junior. "Epanechnikov kernel for incomplete data." *Electronics Letters* TBA.21(2017):TBA.

[14] A, Samuel Asante Gyamerah, P. N. B, and D. I. C . "Probabilistic forecasting of crop yields via quantile random forest and Epanechnikov Kernel function - ScienceDirect", *Agricultural and Forest Meteorology* 280.

[15] Ruben, Jesus, Garcia, et al. "Overestimation and Underestimation Biases in Photon Mapping with Non-Constant Kernels", *IEEE transactions on visualization and computer graphics*, 2014.

[16] Shankar D D, Azhakath A S. "Minor blind feature based Steganalysis for calibrated JPEG images with cross validation and classification using SVM and SVM-PSO", *Multimedia Tools and Applications*, 2020:1-20.

[17] Liu Boning, et al. "An Image Coding Approach Based on Mixture-of-experts Regression Using Epanechnikov Kernel", *ICASSP 2019 - 2019 IEEE International Conference on Acoustics, Speech and Signal Processing (ICASSP) IEEE*, 2019.

[18] Boning Liu, Yan Zhao, Xiaomeng Jiang, Shigang Wang, "Three-dimensional Epanechnikov mixture regression in image coding", *Signal Processing*, Volume 185, 2021, 108090, ISSN 0165-1684, https://doi.org/10.1016/j.sigpro.2021.108090.

[19] Conti C, Soares L D, Nunes P., "Dense Light Field Coding: A Survey"[J]. *IEEE Access*, 2020, PP(99):1-1.

[20] M. B. de Carvalho et al., "A 4D DCT-Based Lenslet Light Field Codec," 2018 *25th IEEE International Conference on Image Processing (ICIP)*, 2018, pp. 435-439, doi: 10.1109/ICIP.2018.8451684.

[21] ISO/IEC JTC1/SC29/WG1, "Verification model software version 2.1 on JPEG Pleno light field coding," Geneva, Switzerland, Tech. Rep. ISO/IEC JTC1/SC29/WG1 N83034, 2019.

[22] Ho-Hyun Kang, Dong-Hak Shin, and Eun-Soo Kim, "Compression scheme of sub-images using Karhunen Loeve transform in three-dimensional integral imaging," *Optics Communications*, vol. 281, no. 14, pp. 3640–3647, Jul. 2008.

[23] C.-L. Chang, X. Zhu, P. Ramanathan, and B. Girod, "Light field compression using disparity-compensated lifting and shape adaptation,"

in *IEEE Transactions on Image Processing*, vol. 15, no. 4, pp. 793–806, Apr. 2006.

[24] Chen, Jie, Junhui Hou, and Lap-Pui Chau. "Light field compression with disparity-guided sparse coding based on structural key views." *IEEE Transactions on Image Processing* 27.1 (2017): 314-324.

[25] Marwah, Kshitij, et al. "Compressive light field photography using overcomplete dictionaries and optimized projections." *ACM Transactions on Graphics (TOG)* 32.4 (2013): 1-12.

[26] L. Li, Z. Li, B. Li, D. Liu and H. Li, "Pseudo-Sequence-Based 2-D Hierarchical Coding Structure for Light-Field Image Compression," in *IEEE Journal of Selected Topics in Signal Processing*, vol. 11, no. 7, pp. 1107-1119, Oct. 2017, doi: 10.1109/JSTSP.2017.2725198.

[27] H. Amirpour, M. Pereira and A. Pinheiro, "High Efficient Snake Order Pseudo-Sequence Based Light Field Image Compression," *2018 Data Compression Conference*, 2018, pp. 397-397, doi: 10.1109/DCC.2018.00050.

[28] S. Zhao, Z. Chen, K. Yang, and H. Huang, "Light field image coding with hybrid scan order," in *2016 Visual Communications and Image Processing (VCIP)*, Nov. 2016, pp. 1–4.

[29] C. Jia et al., "Optimized inter-view prediction based light field image compression with adaptive reconstruction," *2017 IEEE International Conference on Image Processing (ICIP)*, 2017, pp. 4572-4576, doi: 10.1109/ICIP.2017.8297148.

[30] G. Wang, W. Xiang, M. Pickering and C. W. Chen, "Light Field Multi-View Video Coding With Two-Directional Parallel Inter-View Prediction," in *IEEE Transactions on Image Processing*, vol. 25, no. 11, pp. 5104-5117, Nov. 2016, doi: 10.1109/TIP.2016.2603602.

[31] D. Liu, L. Wang, L. Li, Zhiwei Xiong, Feng Wu and Wenjun Zeng, "Pseudo-sequence-based light field image compression," *2016 IEEE International Conference on Multimedia & Expo Workshops (ICMEW)*, 2016, pp. 1-4.

[32] X. Huang, P. An, L. Shen and R. Ma, "Efficient Light Field Images Compression Method Based on Depth Estimation and Optimization," in *IEEE Access*, vol. 6, pp. 48984-48993, 2018, doi: 10.1109/ACCESS.2018.2867862.

[33] P. Astola and I. Tabus, "Coding of Light Fields Using Disparity-Based Sparse Prediction," in *IEEE Access*, vol. 7, pp. 176820-176837, 2019, doi: 10.1109/ACCESS.2019.2957934.

[34] I. Viola, H. Petric Maretic, P. Frossard, and T. Ebrahimi, "A graph learning approach for light field image compression," in *SPI Applications of Digital Image Processing XLI*, vol. 10752, San Diego, CA, US, Sep. 2018, p. 107520E.

[35] E. Dib, M. L. Pendu and C. Guillemot, "Light Field Compression Using Fourier Disparity Layers," *2019 IEEE International Conference on Image Processing (ICIP)*, 2019, pp. 3751-3755, doi: 10.1109/ICIP.2019.8803756.

[36] Z. Zhao, S. Wang, C. Jia, X. Zhang, S. Ma and J. Yang, "Light Field Image Compression Based on Deep Learning," *2018 IEEE International Conference on Multimedia and Expo (ICME)*, 2018, pp. 1-6, doi: 10.1109/ICME.2018.8486546.

[37] C. Jia, X. Zhang, S. Wang, S. Wang and S. Ma, "Light Field Image Compression Using Generative Adversarial Network-Based View Synthesis," in *IEEE Journal on Emerging and Selected Topics in Circuits and Systems*, vol. 9, no. 1, pp. 177-189, March 2019, doi: 10.1109/JETCAS.2018.2886642.

[38] X. Zhang et al., "Surface Light Field Compression Using a Point Cloud Codec," in *IEEE Journal on Emerging and Selected Topics in Circuits and Systems*, vol. 9, no. 1, pp. 163-176, March 2019, doi: 10.1109/JETCAS.2018.2883479.

[39] Yuksel, S. E, J. N. Wilson, and P. D. Gader. "Twenty Years of Mixture of Experts", *Neural Networks and Learning Systems*, IEEE Transactions on 23.8(2012):p.1177-1193.

[40] P. Wang, H. Deng, Y. M. Wang, Y. Liu, Y. Zhang, "Kernel density estimation based Gaussian and non-Gaussian random vibration data induction for high-speed train equipment", *IEEE Access* 8 (2020) 90914–90923.

[41] Kubo Y, Nii M, Muto T, et al. "Artificial humeral head modeling using Kmeans++ clustering and PCA[C]", *2020 IEEE 2nd Global Conference on Life Sciences and Technologies (LifeTech)*. IEEE, 2020.

[42] Moon T. K., "The expectation-maximization algorithm." *Signal Processing Magazine IEEE* 13.6(1996):47-60.

[43] Zhou W, Bovik A C, "Mean squared error: Love it or leave it? A new look at Signal Fidelity Measures", *IEEE Signal Processing Magazine*, 2009, 26(1):98-117.

[44] Guo M, Lyu Y, Wang S, "Real Scene Pickup Method of Elemental Image Array Based on Convergent Camera Array", *IEEE Access*, 2020, PP(99):1-1.

[45] Amirpour Hadi, M. Pereira, and A. Pinheiro, "High Efficient Snake Order Pseudo-Sequence Based Light Field Image Compression", *2018 Data Compression Conference* (2018):397-397.

[46] Sullivan G J, Ohm J R, Han W J, et al. "Overview of the High Efficiency Video Coding (HEVC) Standard", *IEEE Transactions on Circuits & Systems for Video Technology*, 2013, 22(12):1649-1668.

[47] D. Marpe, H. Schwarz, T. Wiegand, "Context-based adaptive binary arithmetic coding in the H.264/AVC video compression standard", *IEEE Transactions on Circuits and Systems for Video Technology* 13 (7) (2003) 620–636.

[48] Wang Y, Liu D, Ma S, et al. "Ensemble Learning-Based Rate-Distortion Optimization for End-to-End Image Compression", *IEEE Transactions on Circuits and Systems for Video Technology*, 2020, PP(99):1-1.

[49] Dabov K, Foi A, Katkovnik V, et al. "Image Denoising by Sparse 3-D Transform-Domain Collaborative Filtering", *IEEE Transactions on Image Processing*, 2007, 16(8):p.2080-2095.

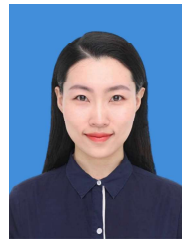

**Boning Liu** received her B.S. degree in communication engineering from Jilin University, China, in 2016. She is currently a Ph.D candidate in Jilin University. Her main research interest is modeling, multimedia signal processing, non-Gaussian signal processing, and image coding.

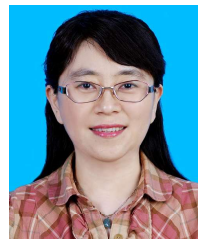

**Yan Zhao** received her B.S. degree in communication engineering in 1993 from Changchun Institute of Posts and Telecommunications, M.S. degree in communication and electronic in 1999 from Jilin University of Technology, and Ph.D.degree in communication and information system in 2003 from Jilin University. She currently is a professor of communication engineering. Her research interests include image and video processing, multimedia signal processing.

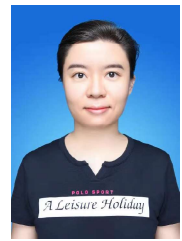

**Xiaomeng Jiang** received her B.S. degree in mathematics and applied mathematics in 2010 from Beijing Normal University and Ph.D degree in fundamental mathematics in 2017 from Jilin University. She currently is a lecturer of finacial mathematics. Her research interests include stochastic analysis and stochastic dynamical systems.

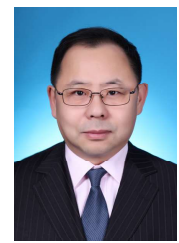

**Shigang Wang** received his B.S. degree in 1983 from Northeastern University, his M.S. degree in communication and electronic in 1998 from Jilin University of Technology, and his Ph.D. degree in communication and information system in 2001from Jilin University. He currently is a professor of communication engineering. His research interests include image and video coding, multidimensional signal processing.

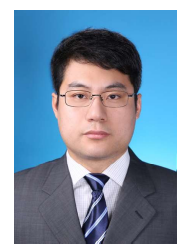

**Jian Wei** received his B.S. degree in communication engineering from Jilin University in 2008, his M.S. degree in communication and information systems from Jilin University in 2011, and his Ph.D degree in informatics from Tuebingen University in 2016. Currently, he is an associate professor in the Department of Communication Engineering, Jilin University. His research interests include multi-view stereo and 3D display.